\documentclass[aps,twocolumn,showpacs,preprintnumbers,floatfix]{revtex4-1}
\usepackage{graphicx}
\usepackage{epsfig}
\usepackage{color}
\usepackage{dcolumn}
\usepackage[colorlinks,linkcolor=red,anchorcolor=green,citecolor=blue,CJKbookmarks=True]{hyperref}
\usepackage{bbm}
\begin{document}

\title{Color Halo Scenario of Charmonium-like Hybrids}

\author{\small
Yunheng Ma${}^{1,2}$,
Wei Sun${}^{1,3}$, 
Ying Chen${}^{1,2}$\thanks{cheny@ihep.ac.cn},
Ming Gong${}^{1,2}$ and 
Zhaofeng Liu${}^{1,2}$
}

\affiliation{\small
$^1$~Institute of High Energy Physics, Chinese Academy of Sciences, Beijing 100049, P.R. China \\
$^2$~School of Physics, University of Chinese Academy of Sciences, Beijing 100049, P.R. China\\
$^3$~Thomas Jefferson National Accelerator Facility, 12000 Jefferson Avenue, Newport News, VA 23606, USA
}

\begin{abstract}
The internal structures of $J^{PC}=1^{--}, (0,1,2)^{-+}$ charmonium-like hybrids are
investigated under lattice QCD in the quenched approximation. We define the Bethe-Salpeter wave
function $\Phi_n(r)$ in the Coulomb gauge as the matrix element of a spatially extended hybrid-
like operator $\bar{c}{c}g$ between the vacuum and $n$-th state for each $J^{PC}$ with $r$
being the spatial separation between a localized $\bar{c}c$ component and the chromomagnetic
strength tensor. These wave functions exhibit some similarities for states with the aforementioned different
quantum numbers, and their $r$-behaviors (no node for the ground states and one node for the
first excited states) imply that $r$ can be a meaningful dynamical variable for these states. 
Additionally, the mass splittings of the ground states and first excited states of charmonium-like 
hybrids in these channels are obtained for the first time to be approximately 1.2-1.4 GeV. 
These results do not support the flux-tube description of heavy-quarkonium-like hybrids in 
the Born-Oppenheimer approximation. In contrast, a charmonium-like hybrid can be viewed as a ``color halo" 
charmonium for which a relatively localized color octet $\bar{c}c$ is surrounded by gluonic degrees of freedom, 
which can readily decay into a charmonium state along with one or more light hadrons. 
The color halo picture is compatible with the decay properties of $Y(4260)$ and suggests LHCb and BelleII 
to search for $(0,1,2)^{-+}$ charmonium-like hybrids in $\chi_{c0,1,2}\eta$ and $J/\psi \omega (\phi)$ final states.
\end{abstract}

 \maketitle

\section{Introduction}
In the constituent quark model, mesons are interpreted as quark-antiquark ($q\bar{q}$) bound states.
However, since gluons are also fundamental degrees of freedom in quantum chromodynamics (QCD), they are expected
to build hadrons either by themselves, such as glueballs, or make create hybrids with quarks.
Glueballs and hybrids are exotic hadron states that have been searched for a long time through
experiments. For the hybrid mesons (denoted by $Q\bar{Q}g$) involving a heavy quark-antiquark pair $Q\bar{Q}$,
an interesting phenomenological description is the flux-tube picture in the leading Born-Oppenheimer approximation~\cite{Juge:1997nc,Juge:1999ie},
in which the gluonic excitations are considered as fast degrees of freedom, which concentrate along the $Q\bar{Q}$ axis and result in an instantaneous confining potential that obeys the cylindrical symmetry along the $Q\bar{Q}$ axis and the refection symmetry with respect to the $Q\bar{Q}$ midpoint. These gluonic excitations provide effective potentials between the $Q\bar{Q}$ pair, which are frequently labeled as $\Lambda_\eta^\epsilon$ in analogy with diatomic molecules, where $\Lambda=0,1,2,\cdots$ is the magnitude of the angular momentum of the gluon state projecting to the axis and is conventionally denoted by $\Sigma, \Pi, \Delta$, respectively; $\eta=\pm $ is the $CP$ quantum number of the gluon state, and $\epsilon=\pm$ is the reflection quantum number of the gluon state with respect to a plane perpendicular to the axis at the midpoint of the axis. Based on this description, many phenomenological studies have been conducted on $Q\bar{Q}g$ hybrids through gluon-excitation pictures\cite{Szczepaniak_2006,Szczepaniak_2006_2} or by solving the non-relativistic Schr\"{o}dinger equations of $Q\bar{Q}$ systems with this type of potential\cite{Braaten:2014qka,Akbar:2020vuo}.

$Q\bar{Q}g$ hybrids, particularly the charmonium-like hybrids $c\bar{c}g$, have also been extensively investigated in lattice QCD \cite{Lacock:1996vy,Bernard:1997ib,Liao:2002rj,Bernard:2003jd,Mei:2002ip,Dudek:2009qf, Yang:2012gz,Dudek:2008sz,Dudek:2009kk}.
A recent lattice calculation~\cite{Liu:2012ze} demonstrated the existence of a $\{1^{--},(0,1,2)^{-+}\}$ charmonium-like supermultiplet with nearly degenerate masses of approximately 4.2-4.4 GeV, which overlaps strongly to $c\bar{c}g$ type operators. These states may have similar internal dynamics, while the different quantum numbers are due to the different couplings of the spin states of the $\bar{c}c$ and chromomagnetic excitation ($J^P=1^+$). In the Born-Oppenheimer potential picture, this supermultiplet can be assigned to $\Pi_u^+(1P)$ states since they are the lowest in mass. The properties of this supermultiplet have practical interest for current experimental and theoretical studies of $XYZ$ particles (see Ref.~\cite{Brambilla:2019esw} for a review). Among the $XYZ$ particles, $Y(4260)$ (or $\psi(4230)$ as named by
PDG 2018~\cite{Tanabashi:2018oca}) can have the possible assignment of a vector charmonium-like hybrid~\cite{Zhu:2005hp} and can be a member of the aforementioned supermultiplet, owing to its strange production and decay properties, as well as its mass adjacent to that of the $1^{-+}$ charmonium-like state predicted by previous phenomenological and lattice QCD studies. In addition to its mass, a quenched lattice calculation also predicted the leptonic decay width of the vector charmonium-like hybrid to be approximately smaller than 40 eV~\cite{Chen:2016ejo}, which explains to some extent the absence of $Y(4260)$ (if a hybrid) in the $R$-value scan of $e^+e^-$ annihilation processes; this was compatible with the estimate from its isospin symmetric decays~\cite{Gao:2017sqa,Yuan:2018inv}. Therefore, a joint study of this multiplet is necessary to understand the experimental observations relevant to $Y(4260)$ and predict the properties of other members to be searched, particularly the state of the exotic $1^{-+}$ quantum number.

In this work, we investigated the internal structure of the $\{1^{--},(0,1,2)^{-+}\}$ charmonium-like hybrid supermultiplet under lattice QCD. As an exploratory study, we adopt the quenched approximation, in which at least the $1^{-+}$ hybrid is well-defined. By constructing spatially extended operators for hybrids, we extracted their Bethe-Salpeter (BS) wave functions through the corresponding correlation functions calculated in the Coulomb gauge.

\section{Formalism} \label{sec:fromalism}
We generated gauge configurations on two anisotropic
lattices using the tadpole improved gauge action~\cite{Morningstar:1999rf, Chen:2005mg}. The aspect ratio is set to be $\xi=a_s/a_t=5$, where $a_s$ and $a_t$ are the spatial and temporal lattice spacings, respectively. The much finer lattice in the temporal direction
enables us to address heavy particles on relatively coarse lattices. The configuration parameters are listed in Table~\ref{tab:lattice}, where the values of $a_s$
are determined from $r_0^{-1}=410(20)$ MeV. For the charm quark, we use
the tadpole improved clover action for anisotropic lattices ~\cite{Su:2004sc}, and the bare charm quark mass is determined by the physical mass of $J/\psi$, $m_{J/\psi}=3.097$ GeV. As will be addressed in the following sections, we use spatially extended operators to calculate the relevant correlation functions; therefore, the configurations are first
fixed to the Coulomb gauge through the standard gauge fixing procedure~\cite{Giusti:2001xf} in lattice QCD studies before the quark propagators are computed.
\begin{table}[t]
	\centering \caption{\label{tab:lattice}
		Input parameters for the calculation. Values for the coupling $\beta$, anisotropy
		$\xi$, lattice spacing $a_s$, lattice size, and number of measurements are
		listed.}
	\begin{ruledtabular}
		\begin{tabular}{cccccc}
			$\beta$ &  $\xi$  & $a_s$(fm) & $La_s$(fm) & $L^3\times T$ & $N_{conf}$ \\\hline
			2.4  & 5 & 0.222(2) & 3.55 &$16^3\times 160$ & 500 \\
			2.8  & 5 & 0.138(1) & 3.31 &$24^3\times 192$ & 200 \\
		\end{tabular}
	\end{ruledtabular}
\end{table}

In this study, we extracted the spectrum and the BS wave functions of charmonium-like $c\bar{c}g$ hybrids of quantum numbers $\{1^{--},(0,1,2)^{-+}\}$. For a particular quantum number, the state-of-art technique in extracting excited states in lattice QCD is the variational method (VM) based on an operator set $\{O_\alpha(t), \alpha=1,2,\ldots, M\}$ of $M$ different operators. The procedure of VM is outlined as follows: First, the correlation matrix $C_{\alpha\beta}(t)=\langle O_\alpha(t)O_\beta^\dagger(0)\rangle$ is calculated on a gauge ensemble. Second, the generalized eigenvalue problem
\begin{equation}
C_{\alpha\beta}(t+t_D)v_\beta^{(m)}=\lambda^{(m)}(t)C_{\alpha\beta}(t_D)v_\beta^{(m)}
\end{equation}
can be solved to derive the eigenvector $v^{(m)}_\alpha$ corresponding to the $m$-th eigenvalue $\lambda^{(m)}(t)$. We can obtain an optimized operator $O^{(m)}(t)=v^{(m)}_\alpha O_\alpha(t)$ that couples mostly to the $m$-th state of the given quantum number. The key aspect of VM is that the operators $O_\alpha$ in the operator set are as different as possible from each other in the sense that they couple to each state in the spectrum differently, such that the linear combination $O^{(m)}(t)=v^{(m)}_\alpha O_\alpha(t)$ produces almost distinct operators for different $m$ values. Therefore, smeared quark and gluon fields are conventionally adopted in constructing operators $O_\alpha$ using the conjecture that the operators with different smearing sizes satisfy the requirement. VM has been extensively and successfully applied in the precise determination of energy levels in lattice QCD studies. However, VM is not suitable for the derivation of hadron decay constants and form factors, which are usually defined through the matrix elements of
local operators between physical states.

The BS wave function is defined by the matrix element of a spatially extended operator $O(r)$ between the vacuum and a hadron state, i.e., $\Phi_n(r)\sim\langle \Omega|O(r)|n \rangle$, where $r$ reflects the internal spatial
coordinates of a hadron. The effect of $O(r)$ on the vacuum state $|\Omega\rangle$ is
\begin{equation}
O^\dagger(r)|\Omega\rangle=\sum\limits_n \frac{1}{2m_nV}|n\rangle\langle n|O^\dagger(r)|\Omega\rangle \sim \sum\limits_n \Phi_n^*(r)|n\rangle,
\end{equation}
which implies that the BS wave function $\Phi_n(r)$ actually controls the coupling of the operator $O(r)$ to the state $|n\rangle$. Therefore, if the $r$ behaviors of $\Phi_n(r)$ are different for different state $|n\rangle$, then $O(r)$ can be considered a different operator for different $r$ values. On the other hand, the correlation function of $O(r,t)$ and some other source operator $O_S(0)$ has the following spectral expression
\begin{eqnarray}\label{form}
C(r,t)&=&\langle \Omega|O(r,t)O^\dagger_S(0)|\Omega\rangle\nonumber\\
      &=&\sum\limits_n \frac{1}{2m_nV}\langle \Omega|O(r)|n\rangle\langle n|O_S^\dagger|\Omega\rangle e^{-m_n t}\nonumber\\
      &\equiv&\sum\limits_n \Phi_n(r)e^{-m_n t}
\end{eqnarray}
where $m_n$ is the mass of the $n$-th state and the $r$-independent factors are
absorbed into $\Phi_n(r)$ in the last equality. Since the $m_n$ spectrum is common for different $r$ values, if $N_r$ correlation functions $C(r,t)$ with different $r$ values are calculated, then in proper time windows $t\in [t_\mathrm{min},t_\mathrm{max}]$, the $m_n$ and $\Phi_n(r)$ values for $n=1,2,\ldots, N_m$ can be determined by fitting these correlation functions simultaneously with the function form in Eq.~(\ref{form}) involving $N_m$ mass terms. Specifically, if the time window $[t_\mathrm{min},t_\mathrm{max}]$ is uniform for all values of $r$ and there are $N_t$ time points in this window, there will be $N_r\times N_t$ data points for the $N_m+N_r\times N_m$ parameters to be fitted. Thus, the number of the degrees of freedom can be sufficiently large if $N_t$ is significantly larger than $N_m$.

We adopted the above strategy to study the spectrum and BS wave functions of $\{1^{--},(0,1,2)^{-+}\}$ charmonium-like hybrids. The operator prototypes for these states are selected to be the traditional $\bar{c}\Gamma c \circ B$ type:
\begin{eqnarray}\label{operator}
O_{0^{-+}_H}  (t)&=&\sum\limits_{\mathbf{x}} \bar{c}^a(\mathbf{x},t)\gamma_i
c^b(\mathbf{x},t) B_i^{ab}(\mathbf{x},t),\nonumber\\
O_{1^{-+}_H}^k(t)&=&\sum\limits_{\mathbf{x}} \bar{c}^a(\mathbf{x},t)\gamma_i
c^b(\mathbf{x},t) B_j^{ab}(\mathbf{x},t)\epsilon_{ijk},\nonumber\\
O_{2^{-+}_H}^k(t)&=&\sum\limits_{\mathbf{x}} \bar{c}^a(\mathbf{x},t)\gamma_i
c^b(\mathbf{x},t) B_j^{ab}(\mathbf{x},t)|\epsilon_{ijk}|,\nonumber\\
O_{1^{--}_H}^k(t)&=&\sum\limits_{\mathbf{x}} \bar{c}^a(\mathbf{x},t)\gamma_5
c^b(\mathbf{x},t) B_k^{ab}(\mathbf{x},t),
\end{eqnarray}
where $a,b=1,2,3$ are color indices, $B_k$ is the chromo-magnetic field strength, and the summation over $\mathbf{x}$ guarantees the operator to couple to a state in its rest frame. We can easily observe that the quark-bilinear operators, $\bar{c}\gamma_i c$ and $\bar{c}\gamma_5 c$, are a spin triplet and singlet, respectively, in the non-relativistic approximation.
Note that the spatial symmetry of the lattice is described by the octahedral group, $O$, whose irreducible representations are $R=A_1, A_2, E, T_1$ and $T_2$, whose dimensions $d(R)$ are 1, 1, 2, 3, 3 respectively. Group $O$ is a subgroup of $SU(2)$ such that the subduced representation of $SU(2)$ with respect to $O$ is generally reducible. Table~\ref{subduced rep} shows
the reduction of the subduced representation of $SU(2)$ up to $J=4$. For instance, the scalar and
pseudoscalar with $J=0$ states are represented by $A_1$, and tensor states with $J=2$ are reduced to
the direct sum of $E$ and $T_2$, i.e. $(J=2)\downarrow O=E\oplus T_2$. Although the operators on the right-hand side of Eq.(\ref{operator}) are the representations $R^{PC}=A_1^{-+}, T_1^{-+}, T_2^{-+}$ and $T_1^{--}$ of the lattice symmetry group, respectively, they have the same forms as the $0^{-+}$, $1^{-+}$, $2^{-+}$ and $1^{--}$ operators in the continuum limit. Therefore, we do not distinguish them from each other under the assumption that the lattice discretization effects are small.
\begin{table}[t]
	\centering \caption{\label{subduced rep} Reduction of subduced representations of $SU(2)$ with respect to octahedral group $O$ up to $J=4$.}
	\begin{ruledtabular}
		\begin{tabular}{cccccc}
		$R$	    &  $A_1$  & $A_2$ & $E$   & $T_1$ & $T_2$ \\
		$d(R)$  &    1    &   1       &  2    &   3   &  3 \\
		$J$     & 0, 4     &   3       & 2, 4, 5 &1, 3, 4  & 2, 3, 4 \\
		\end{tabular}
	\end{ruledtabular}
\end{table}
To investigate the inner structure of the charmonium-like states through BS wave functions, we introduce two types of spatially extended operators based on the operator prototypes in Eq.(\ref{operator}). Type-I operators are constructed by splitting the $\bar{c}\Gamma c$ component from the chromomagnetic field strength operator $\mathbf{B}$ by a spatial separation $r$:
\begin{equation}\label{eq:type-I}
O_R^{(I)}(r,t)=\frac{1}{D_r}\sum\limits_{\mathbf{x},|\mathbf{r}|=r}\bar{c}(\mathbf{x},t)\Gamma_R c(\mathbf{x},t)\circ \mathbf{B}(\mathbf{x}+\mathbf{r},t)
\end{equation}
where $D(r)$ is the number of $\mathbf{r}$ values that satisfy $|\mathbf{r}|=r$, $\Gamma_R$ is the gamma matrix appearing in the operator prototype $O_R$ in Eq.(\ref{operator}) with $R=(0,1,2)^{-+}$ and $1^{--}$, and the symbol ``$\circ$" indicates the proper color summation and specific combination of the spatial indexes. Type-II
operators are obtained by assuming a spatial separation between $\bar{c}$ and $c$ with $\mathbf{B}$ being at the same point as $c$:
\begin{equation}\label{eq:type-II}
O_R^{(II)}(r,t)=\frac{1}{D_r}\sum\limits_{\mathbf{x},|\mathbf{r}|=r}\bar{c}(\mathbf{x+r},t)\Gamma_R c(\mathbf{x},t)\circ \mathbf{B}(\mathbf{x},t)+c.c.
\end{equation}
where ``$c.c.$" is the complex conjugate of the first term that is required to guarantee the correct parity and charge conjugate quantum numbers.

These two types of operators are not gauge invariant; therefore, the correlation functions involving these operators should be calculated in a fixed gauge. In practice, we work in the Coulomb gauge and calculate the following correlation functions for each quantum number $R$:
\begin{equation}\label{two-point}
C_R(r,t)=\langle O_R(r,t+\tau) O_R^{(W)\dagger}(\tau)\rangle
\end{equation}
where $O_R^{(W)}(\tau)$ is a wall-source operator defined on the time-slice $\tau$ for $R$ as
\begin{equation}\label{wall-src}
O_R^{(W)}(\tau)=\sum\limits_{\mathbf{y,z}}\bar{c}(\mathbf{y},\tau)\Gamma_R c(\mathbf{z},\tau) \circ\mathbf{B}(\mathbf{z},\tau)+c.c.
\end{equation}
In the data analysis stage, these correlation functions are parameterized to the function form of Eq.(\ref{form}), based on which masses $m_n$ and BS wave functions $\Phi_n(r)$ are derived simultaneously through a correlated joint fit to the lattice data of $C_R(r,t)$ at different values of $r$ and $t$.
\section{Results using type-I operators}
In this section, we present the results of the spectrum and the BS wave functions of $\{1^{--},(0,1,2)^{-+}\}$ states using the type-I operators $O_R^{(I)}(r,t)$, where $r$ is the spatial separation between the $c\bar{c}$ and $\mathbf{B}$ components. In each channel, the Coulomb wall-source correlation functions $C_R(r,t)$ are modeled as
\begin{eqnarray}\label{model}
C_R(r,t)&\equiv& \frac{1}{N_\tau}\sum\limits_{\tau}\langle O_R^{(I)}(r,t+\tau)O_R^{(W)\dagger}(\tau)\rangle\nonumber\\
&=& \sum\limits_n \Phi_n^{(I)}(r)e^{-m_n t},
\end{eqnarray}
where $\tau$ is the source time slice and $N_\tau=10$ is the number of sources on each configuration, and the spatial indexes of operators are summed implicitly in $1^{--}$ and  $(0,1,2)^{-+}$ channels. For each quantum number, since
$m_n$ is the same for different $r$ values, we perform simultaneous multi-exponential fits to
$C_R(r,t)$ using a correlated minimal-$\chi^2$ fit method with the jackknife covariance
matrix, which results in the parameters $m_n$ and $\Phi_n^{(I)}(r)$ directly.
\begin{table}[t]
	\caption{\label{tab:1-+mass}
		Fitted masses $m_n,n=1,2,3$ of $1^{-+}$ states at different $t_\mathrm{min}$ and the $\chi^2$ per degree of freedom of each fit. The mass values are converted to physical units using the lattice spacings listed in Tab.~\ref{tab:lattice}. }
	\begin{ruledtabular}
		\begin{tabular}{cccccc}
			&$t_{\rm min}/a_t$ & $\chi^2/{\rm dof}$ & $m_1$ (GeV) & $m_2$ (GeV) & $m_3$ (GeV)  \\\hline
			$\beta=2.4$&	5   &  0.94     &  4.300(12)  &  5.35(18) &  6.79(59)  \\
			&	4   &  0.82     &  4.300(08)  &  5.48(09) &  7.35(35)  \\
			&	3   &  0.79     &  4.297(07)  &  5.42(06) &  6.99(13)  \\
			&	2   &  1.03     &  4.309(05)  &  5.59(03) &  7.35(07)  \\
			&	1   &  2.39     &  4.334(03)  &  5.74(02) &  7.64(03)  \\\hline
			$\beta=2.8$&    8   &  1.17     &  4.270(17)  &  5.72(10) &  7.80(61)  \\
			&    7   &  1.31     &  4.267(14)  &  5.52(09) &  7.70(28)  \\
			&    6   &  1.15     &  4.281(11)  &  5.56(06) &  7.57(18)  \\
			&    5   &  1.37     &  4.289(08)  &  5.66(04) &  7.68(11)  \\
			&    4   &  1.52     &  4.308(07)  &  5.76(03) &  7.84(07)
		\end{tabular}
	\end{ruledtabular}
\end{table}
\begin{table*}[t]
	\caption{\label{tab:1-+wave}
		BS wave functions $\Phi_n^{(I)}(r),n=1,2,3$ of $1^{-+}$ states at $t_\mathrm{min}/a_t=2$ for $\beta=2.4$ and $t_\mathrm{min}/a_t=5$ for $\beta=2.8$. }
	\begin{ruledtabular}
		\begin{tabular}{rrrr|rrrr}
			& $\beta=2.4$&             &              &     & $\beta=2.8$  &             &\\
			$r/a_s$ & $\Phi_1^{(I)}(r)$ & $\Phi_2^{(I)}(r)$ & $\Phi_3^{(I)}(r)$ & $r/a_s$ & $\Phi_1^{(I)}(r)$ & $\Phi_2^{(I)}(r)$ & $\Phi_3^{(I)}(r)$ \\
			\hline
			0.000  &41.71(54) & 47.51(95) & 15.0(1.3)    & 0.000 & 44.79(1.06) &  104.0(3.6)   &  72.9(5.1)\\
			1.000  &23.62(28) & 17.32(37) &-2.03(42)     & 1.000 & 33.51(77)  &   65.7(2.0)   &  22.3(3.4)\\
			1.414  &13.97(15) &  3.41(28) &-3.66(21)     & 1.414 & 25.73(56)  &   40.2(1.2)   &  -3.1(2.3)\\
			1.732  & 8.41(08) & -2.31(25) &-1.42(24)     & 1.732 & 20.08(43)  &   23.0(0.8)   & -14.1(1.6)\\
			2.000  & 4.92(04) & -4.49(22) &-1.28(26)     & 2.000 & 16.03(33)  &   12.1(0.7)   & -15.4(1.2)\\
			2.236  & 2.78(02) & -4.83(17) & 0.35(22)     & 2.236 & 12.68(25)  &    3.9(0.7)   & -15.5(1.1)\\
			2.449  & 1.46(02) & -4.13(13) & 1.43(18)     & 2.449 & 10.10(19)  &   -1.5(0.7)   & -13.0(1.0)\\
			2.828  & 0.03(01) & -2.88(08) & 1.32(12)     & 3.000 &  5.10(09)  &   -7.9(0.6)   &  -3.6(1.0)\\
			3.000  &-0.31(01) & -2.03(05) & 1.28(09)     & 3.317 &  3.15(05)  &   -8.5(0.5)   &   1.2(0.9)\\
			3.162  &-0.58(01) & -1.75(06) & 1.08(09)     & 3.606 &  1.82(03)  &   -8.0(0.4)   &   3.9(0.8)\\
			3.317  &-0.65(01) & -1.09(04) & 0.97(06)     & 4.123 &  0.28(02)  &   -5.5(0.2)   &   4.8(0.6)\\
			3.464  &-0.63(01) & -0.43(05) & 0.61(08)     & 4.583 & -0.38(02)  &   -3.5(0.1)   &   4.6(0.4)\\
			3.606  &-0.68(01) & -0.49(04) & 0.58(06)     & 4.899 & -0.59(02)  &   -2.3(0.1)   &   3.8(0.3)\\		
			3.742  &-0.65(01) & -0.15(03) & 0.40(05)     & 5.385 & -0.72(02)  &   -0.8(0.1)   &   1.9(0.2)\\
			4.000  &-0.70(02) & -0.05(07) & 0.09(11)     &       &            &               &
			
		\end{tabular}
	\end{ruledtabular}
\end{table*}
\subsection{$1^{-+}$ states and $2^{-+}$ states}
\begin{figure}[t!]
	\includegraphics[height=5.5cm]{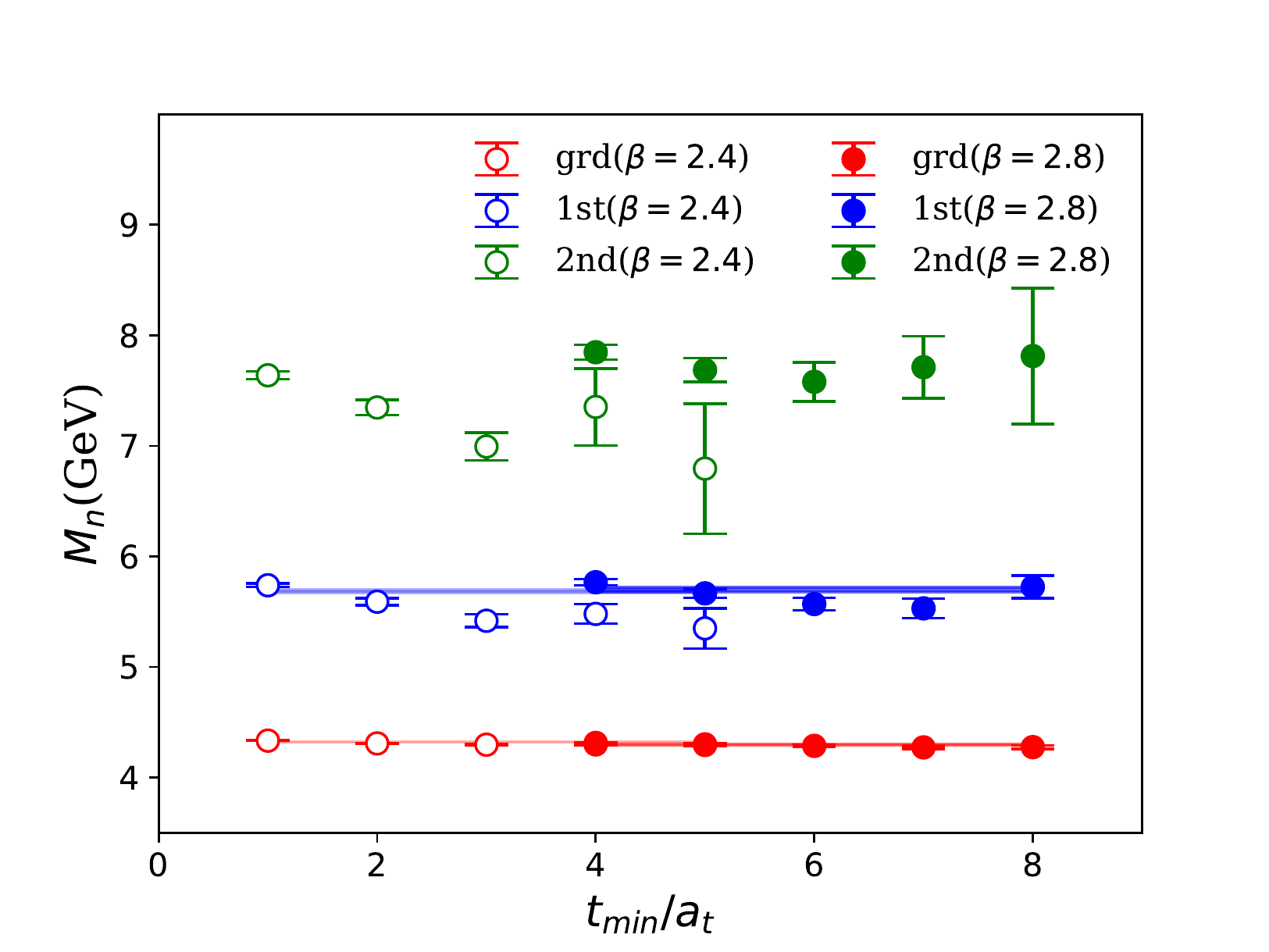}\quad
	\includegraphics[height=5.5cm]{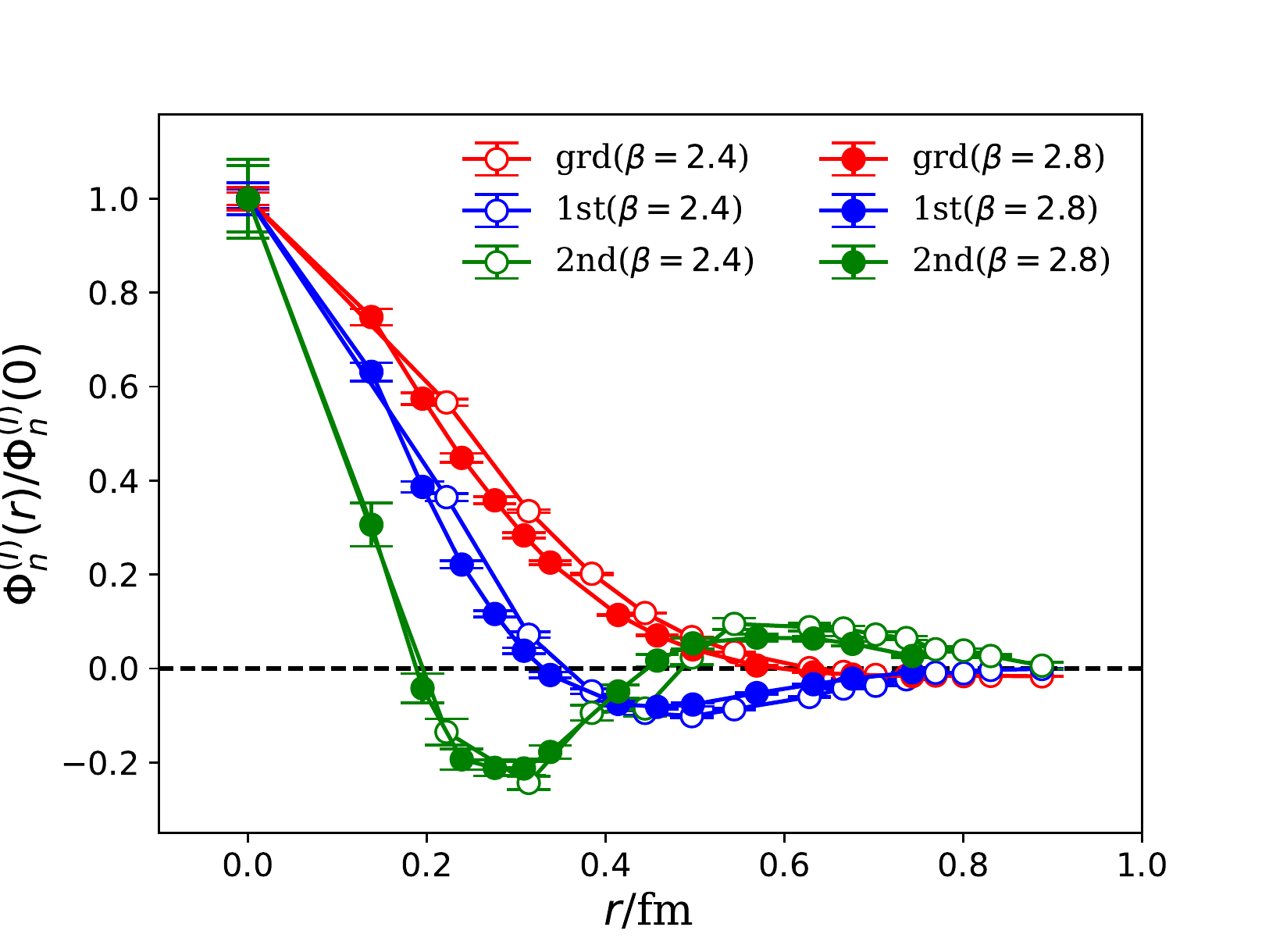}\quad
	\caption{\label{fig:1-+res}
		Mass spectrum and BS wave functions $\Phi_n^{(I)}(r)$ of $1^{-+}$ states. The upper panel
		shows the masses of the lowest three states fitted at different values of $t_{\rm min}$. The
		lower panel shows $\Phi_n^{(I)}(r)$ of these states (normalized by $\Phi_n^{(I)}(0)$). The open and filled circles are the
		results at $t_\mathrm{min}/a_t=2(\beta=2.4)$ and $t_\mathrm{min}/a_t=5(\beta=2.8)$, respectively. We can observe that the finite lattice spacing effect is small. The values of $m_n$ and $r$ are converted to physical units using the lattice spacings listed in Tab.~\ref{tab:lattice}. }
\end{figure}

Since $J^{PC}=1^{-+}$ is prohibited for $q\bar{q}$ mesons, the $1^{-+}$ hidden-charm states must be hybrid-like states in the quenched approximation. Based on the aforementioned numerical strategy, we performed the data fitting using the model in Eq.~(\ref{model}) to $C_R(r,t)$ functions with $r\in[0,r_{\rm max}]$
($n=3, r_{\rm max}=0.89$ fm for $\beta=2.4$ and $n=3, r_{\rm max}=0.74 $ fm for $\beta=2.8$). In the simultaneous fitting procedure, the correlations of data points at different $r$ and $t$ values were considered by calculating their jackknife covariance matrix. The key point is that the sink operator with different $r$ values couples differently to different states; thus, it has different spectral weights $\Phi_n^{(I)}(r)$. We fixed the upper limit of the fitting window ($t_{\rm max}/a_t=19$ for both $\beta=2.4$ and $\beta=2.8$) and let the lower bound $t_{\rm min}$ vary. The fit results of $m_n$ at different $t_{\mathrm{min}}$ values on the two lattices are listed in Table~\ref{tab:1-+mass}, where the masses are converted to the values in physical units using the lattice spacings in Table~\ref{tab:lattice}; the table also lists the $\chi^2$ values per degree of freedom ($\chi^2/\mathrm{dof}$). As shown, the fitted masses are very stable and insensitive to $t_{\rm min}$. Additionally, the masses of the lowest two states exhibit slight finite $a_s$ effects. Figure~\ref{fig:1-+res} shows the fitted masses $m_n$ (in physical units) of the lowest three states versus $t_{\rm min}$, where the colored bands are the averaged mass values weighted with the reciprocals of the error squared at different values of $t_{\rm min}$.
\begin{figure}[t!]
	\includegraphics[height=5.5cm]{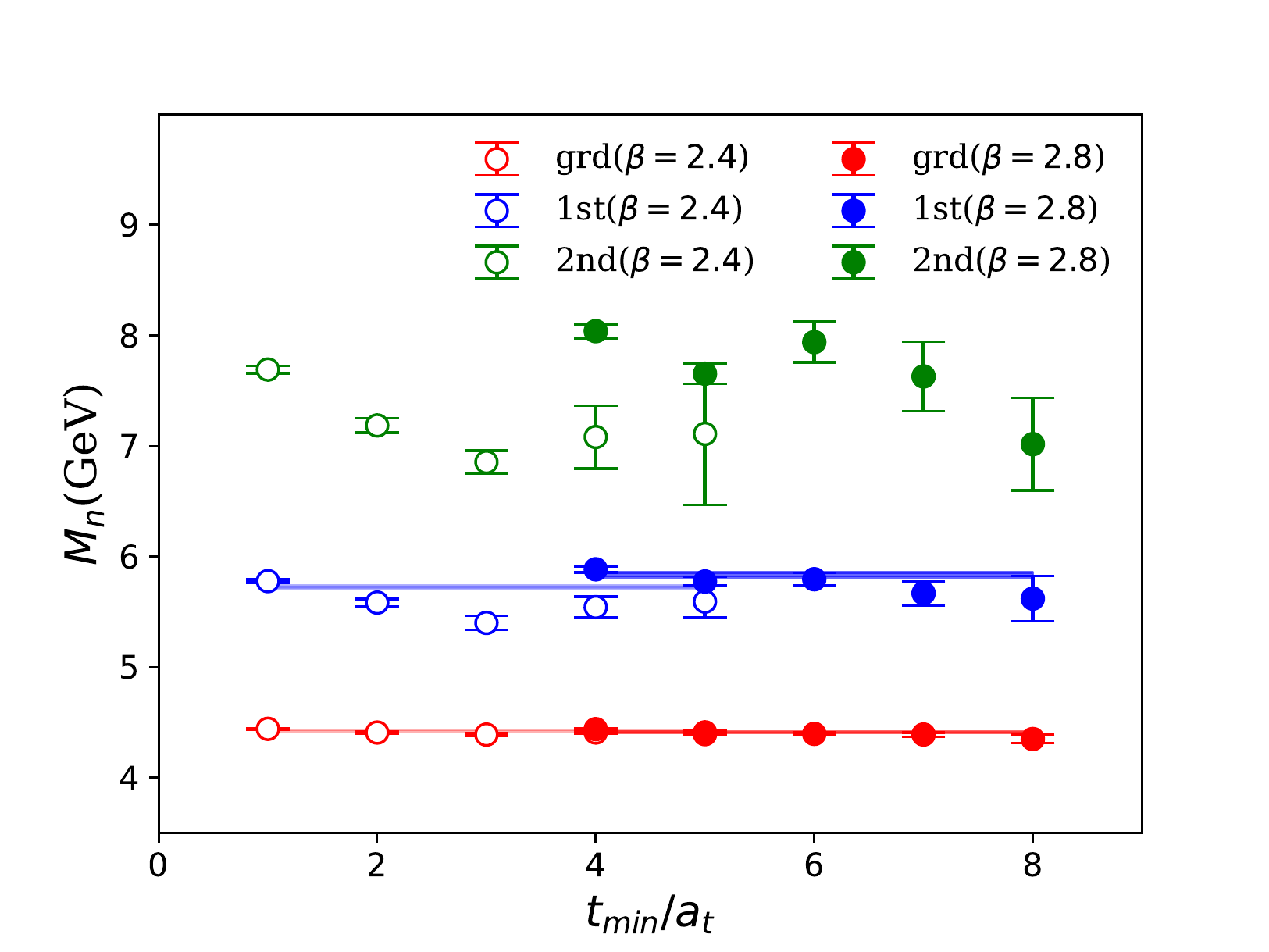}
	\includegraphics[height=5.5cm]{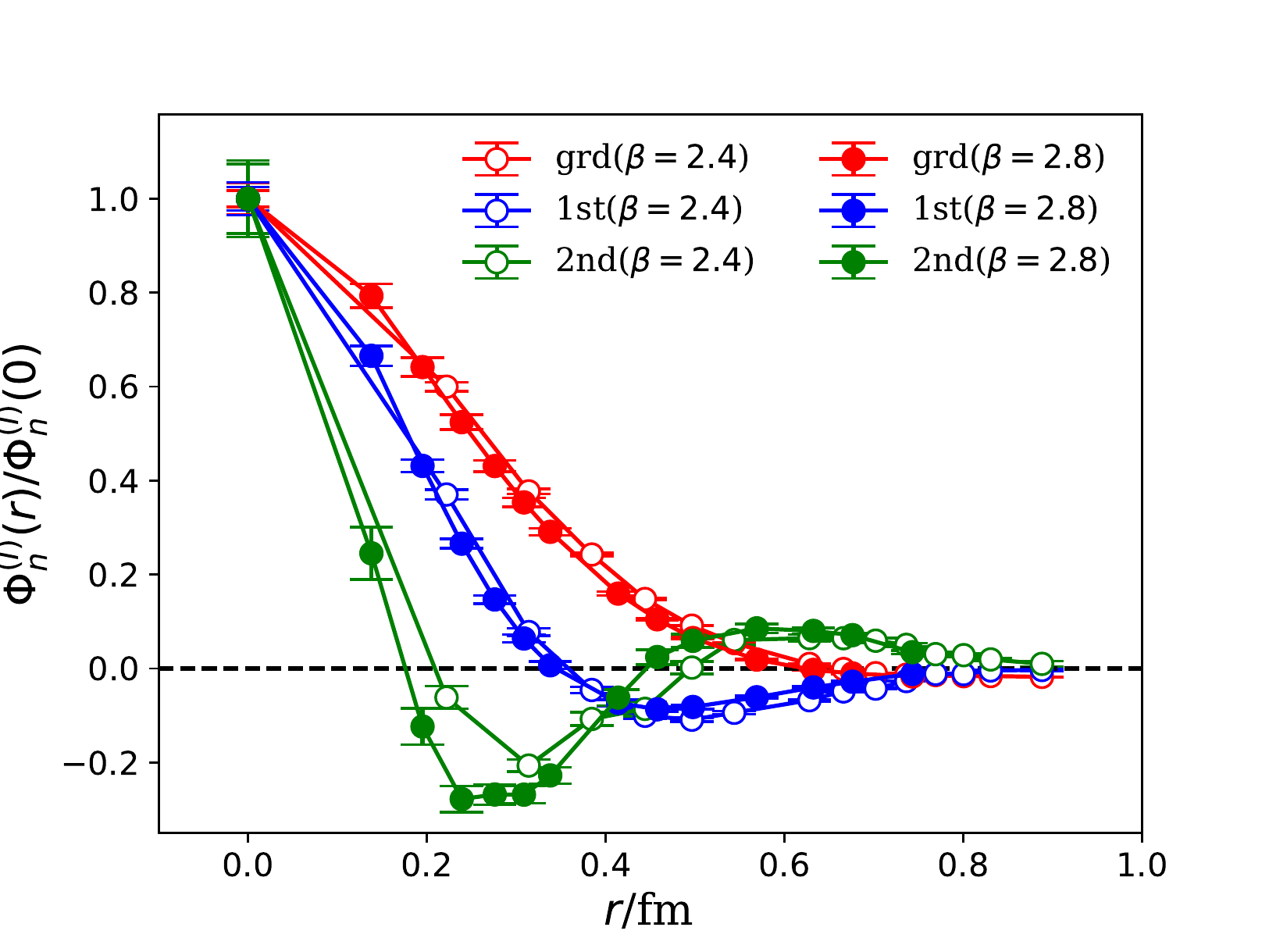}
	\caption{\label{fig:2-+res} Spectrum (upper panel) and BS wave functions (lower panel) of the $2^{-+}$ states. The layout is similar to Fig.~\ref{fig:1-+res}.}
\end{figure}

The $\Phi_n^{(I)}(r)$ functions can be derived simultaneously for the three lowest states. Table~\ref{tab:1-+wave} shows the results at $t_{\rm min}/a_t=2$ for $\beta=2.4$ and $t_{\rm min}/a_t=5$ for $\beta=2.8$, where the fitted parameters have smaller statistical errors and acceptable values of $\chi^2/\mathrm{dof}$ (see Table~\ref{tab:1-+mass}). After the normalization using $\Phi_n^{(I)}(0)$, $\Phi_n^{(I)}(r)$ on the two lattices are shown in the lower panel of Fig.~\ref{fig:1-+res}, where $r$ is in physical units (converted from $a_s$ listed in Tab.~\ref{tab:lattice}). The data points are connected by straight lines to guide the eyes. The radial behaviors of $\Phi_n^{(I)}(r)$ are very clear: $\Phi_1^{(I)}(r)$ has no radial node and $\Phi_2^{(I)}(r)$ has one node and $\Phi_3^{(I)}(r)$ has two nodes.

\begin{table}[t]
 	\caption{\label{tab:2-+mass}
 		Fitted masses $m_n,n=1,2,3$ of $2^{-+}$ states at different $t_\mathrm{min}$ and $\chi^2$ per degree of freedom of each fit. The mass values are converted to the physical units using the lattice spacings listed in Tab.~\ref{tab:lattice}. }
 	\begin{ruledtabular}
 		\begin{tabular}{cccccc}
 			&$t_{\rm min}/a_t$ & $\chi^2/{\rm dof}$ & $m_1$ (GeV) & $m_2$ (GeV) & $m_3$ (GeV)  \\\hline
 			$\beta=2.4$
 			&	5   &  0.86     &  4.408(13)  &  5.59(15) &  7.11(64)  \\
 			&	4   &  0.88     &  4.407(10)  &  5.54(10) &  7.08(28)  \\
 			&	3   &  0.92     &  4.388(10)  &  5.40(06) &  6.85(10)  \\
 			&	2   &  1.04     &  4.407(06)  &  5.58(03) &  7.34(06)  \\
 			&	1   &  2.23     &  4.441(04)  &  5.78(01) &  7.64(03)  \\\hline
 			$\beta=2.8$
 			&   8   &  1.64     &  4.343(36)  &  5.61(21) &  7.01(42)  \\
 			&   7   &  1.75     &  4.384(19)  &  5.66(11) &  7.62(31)  \\
 			&   6   &  1.48     &  4.389(12)  &  5.79(06) &  7.93(18)  \\
 			&   5   &  1.81     &  4.387(11)  &  5.77(04) &  7.64(09)  \\
 			&   4   &  1.86     &  4.432(08)  &  5.88(03) &  8.03(06)
 		\end{tabular}
 	\end{ruledtabular}
\end{table}

$J^{PC}=2^{-+}$ is permitted for conventional $q\bar{q}$ mesons. The mass of the lowest $2^{-+}$
charmonium (termed $\eta_{c2}$) is estimated to be approximately $3.8$ GeV, since it belongs to the $1D$ multiplet $\{2^{-+},(1,2,3)^{--}\}$ whose $1^{--}$ member is $\psi(3770)$ and $2^{--}$ member  $\psi_2(3823)$ was observed by Belle~\cite{Bhardwaj:2013rmw} and BESIII~\cite{Ablikim:2015dlj}. Recently, LHCb also observed a candidate $\psi_3(3842)$ for the $3^{--}$ charmonium~\cite{Aaij:2019evc}. Their masses comply with the prediction of lattice QCD studies~\cite{Liu:2012ze}. In addition to $\eta_{c2}$, lattice QCD studies find a higher $2^{-+}$ state with a mass around $4.4$ GeV, which couples almost exclusively to $c\bar{c}g$ operators~\cite{Liu:2012ze,Yang:2012mya}. This is also the case in this paper. The $2^{-+}$ correlation functions $C_R(r,t)$ using the operators $O_{2^{-+}_H}^k(r,t)$ in Eq.~(\ref{operator}), and the data analysis is the same as for $1^{-+}$. The mass spectrum and BS wave functions (Fig.~\ref{fig:2-+res} and Table~\ref{tab:2-+mass} ) are very similar to those of $1^{-+}$ states.
\begin{table*}[t]
	\caption{\label{tab:P-V-3-24}
		Fitted masses $m_n,n=1,2,3$ of $0^{-+}$ and $1^{--}$ states at different $t_\mathrm{min}$ values on the two lattices, where the mass-values are converted into physical units using the lattice spacings listed in Tab.~\ref{tab:lattice}. The $\chi^2/\mathrm{dof}$ value of each fit is provided to indicate the fitting quality. The $\Phi_n^{(I)}(r)$ values at $r=0$ are also listed.}
	\begin{ruledtabular}
		\begin{tabular}{ccclccccc}
			&$t_{\rm min}/a_t$ & $\chi^2/{\rm dof}$ & $m_1$ (GeV) & $m_2$ (GeV) & $m_3$ (GeV)  & $\Phi_1^{(I)}(0)$  & $\Phi_2^{(I)}(0)$ & $\Phi_3^{(I)}(0)$ \\\hline
					&        &           &             & $\beta=2.4$ & & & &\\\hline
			$0^{-+}$	
			&	13   &  1.19     &  2.985(6)   &  3.75(26) &  4.67(30)
			&  6.87(28)   &  -5.3(3.5)&  37(21) \\
			&	12   &  1.00     &  2.985(4)   &  3.76(20) &  4.61(19)
			&  6.88(22)   &  -5.5(2.8)&  33(10) \\
			&	11   &  0.91     &  2.987(6)   &  3.65(17) &  4.54(12)
			&  7.00(30)   &  -4.5(1.5)&  26(05) \\
			&	10   &  0.83     &  2.985(4)   &  3.78(13) &  4.48(10)
			&  6.90(18)   &  -6.8(2.6)&  29(04) \\
			&    9   &  0.78     &  2.979(3)   &  3.95(09) &  4.62(10)
			&  6.66(10)   &  -7.9(2.4)&  37(03) \\
			&&&&&&&&\\
			$1^{--}$
			&	13   &  1.21     &  3.092(13)   &  3.66(30) &  4.73(26)
			&  6.94(72)    &  -3.8(1.5)&  38(21) \\
			&	12   &  1.07     &  3.092(8)   &  3.82(28) &  4.44(19)
			&  6.88(43)   &  -7.8(7.1)&  30(10) \\
			&	11   &  1.05     &  3.094(12)   &  3.67(25) &  4.43(11)
			&  7.04(67)   &  -5.0(2.2)&  23(05) \\
			&	10   &  1.01     &  3.097(10)   &  3.73(20) &  4.34(10)
			&  7.14(54)   &  -7.6(4.0)&  25(05) \\
			&    9   &  0.86     &  3.082(4)   &  4.09(13) &  4.53(13)
			&  6.44(14)   &  -12.6(8.8)&  39(08) \\\hline
			&        &           &             & $\beta=2.8$ & & & &\\\hline
						$0^{-+}$	
			&    19   & 0.79     &  3.029(14)  &  3.70(36) &  4.45(28)
			&  16.2(1.7)  &  -14(11)  &  32(22) \\
			&    18  &  0.68     &  3.032(15)  &  3.67(30) &  4.41(26)
			&  16.7(1.9)  &  -15(09)  &  33(17) \\
			&    17  &  0.69     &  3.036(15)  &  3.59(20) &  4.55(18)
			&  17.1(2.0)  &  -12(03)  &  34(11) \\
			&    16  &  0.66     &  3.028(17)  &  3.54(22) &  4.71(18)
			&  16.4(2.1)  &  -9(01)    &  30(10) \\
			&    15  &  0.70     &  3.006(03)  &  4.10(09) &  5.07(28)
			&  14.2(2)    &  -14(05)   &  70(19)\\
				&&&&&&&&\\
			$1^{--}$
			&    19   & 0.73     &  3.118(16)  &  3.73(34) &  4.64(33)
			&  14.5(1.8)  &  -9.8(5.0)&  34(22) \\
			&    18  &  0.66     &  3.125(15)  &  3.63(19) &  4.92(28)
			&  15.3(1.9)  &  -8.7(1.8)&  55(27) \\
			&    17  &  0.62     &  3.119(12)  &  3.71(18) &  4.82(22)
			&  14.6(1.2)  &  -8.9(1.9)&  46(15) \\
			&    16  &  0.66     &  3.116(19)  &  3.67(25) &  4.70(19)
			&  14.5(2.0)  &  -7.9(1.5)&  33(10) \\
			&    15  &  0.68     &  3.093(03)  &  4.44(21) &  4.92(48)
			&  12.5(0.2)  &  -26(50)  &  75(33)
		\end{tabular}
	\end{ruledtabular}
\end{table*}
\subsection{$0^{-+}$ and $1^{--}$ states}
\begin{figure*}[t!]
	\includegraphics[height=6cm]{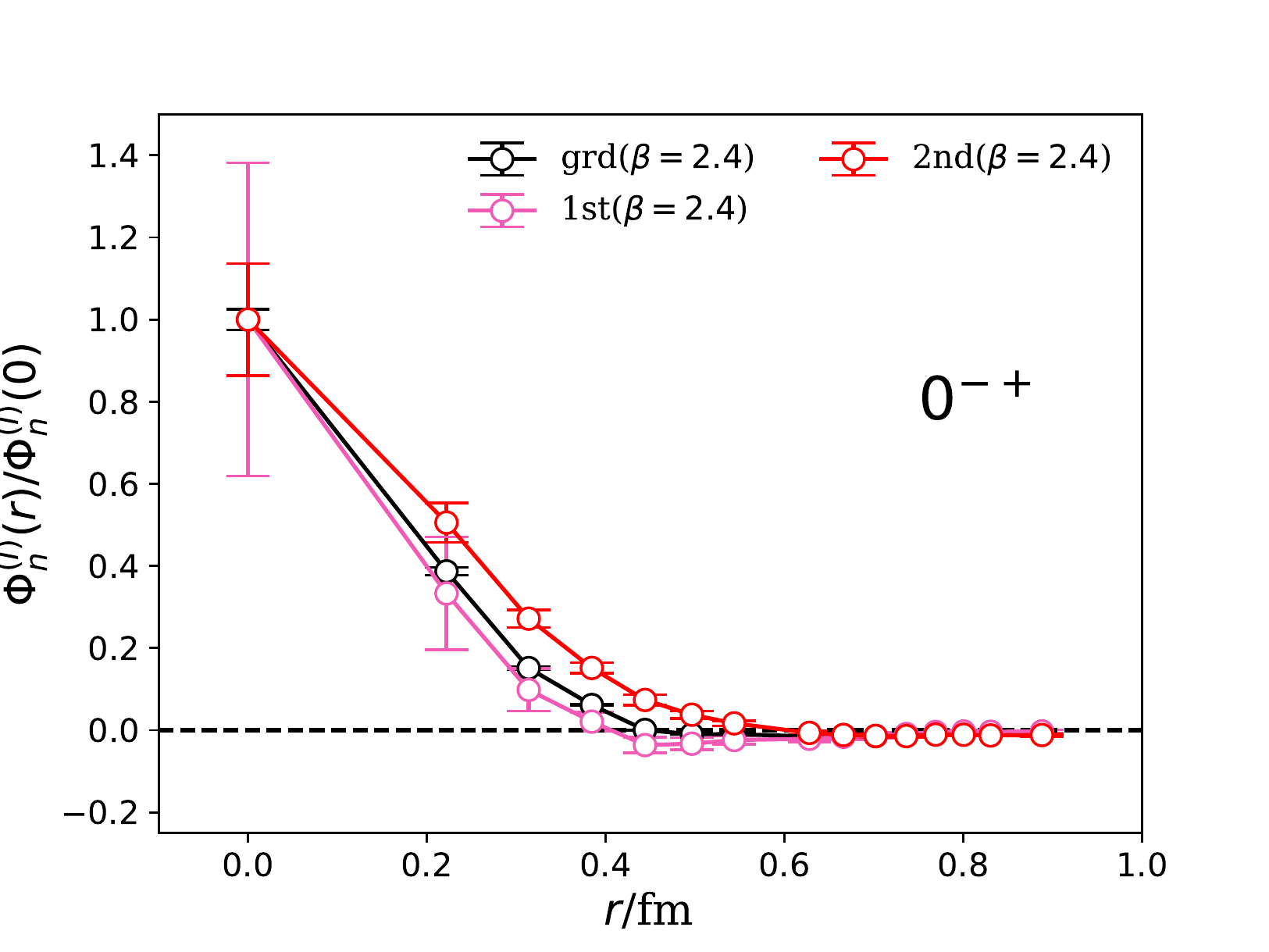}
	\includegraphics[height=6cm]{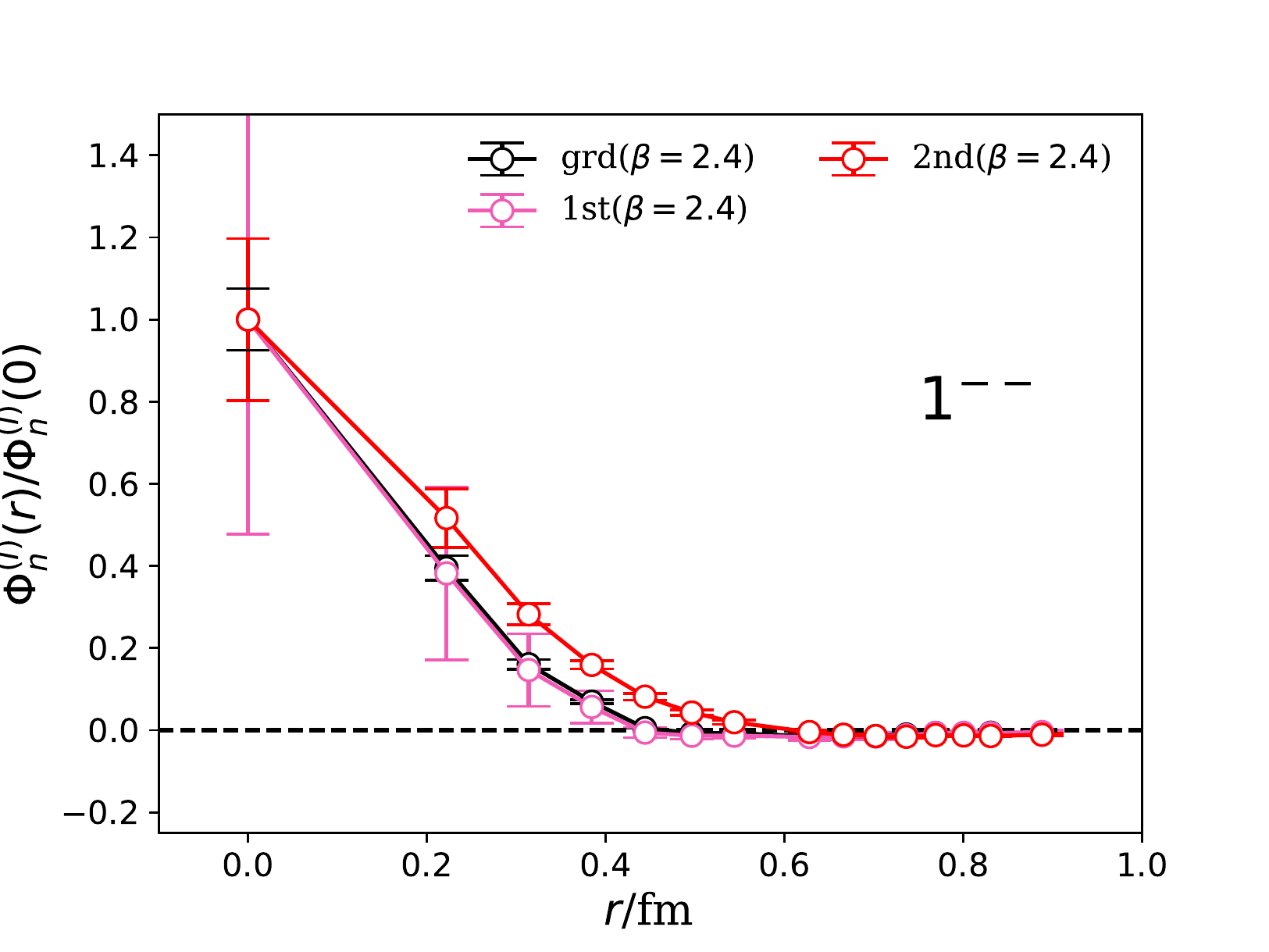}
	\includegraphics[height=6cm]{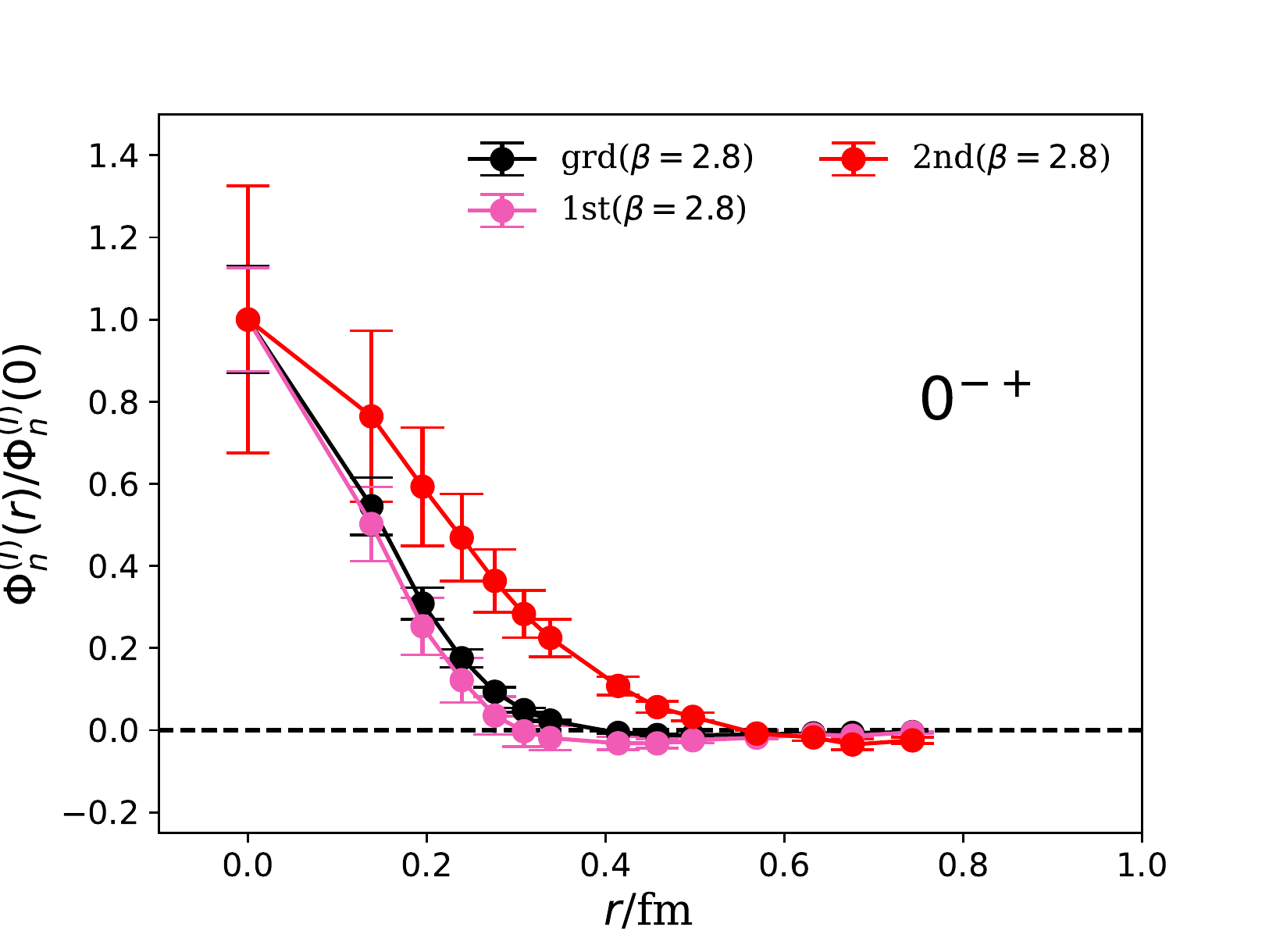}
	\includegraphics[height=6cm]{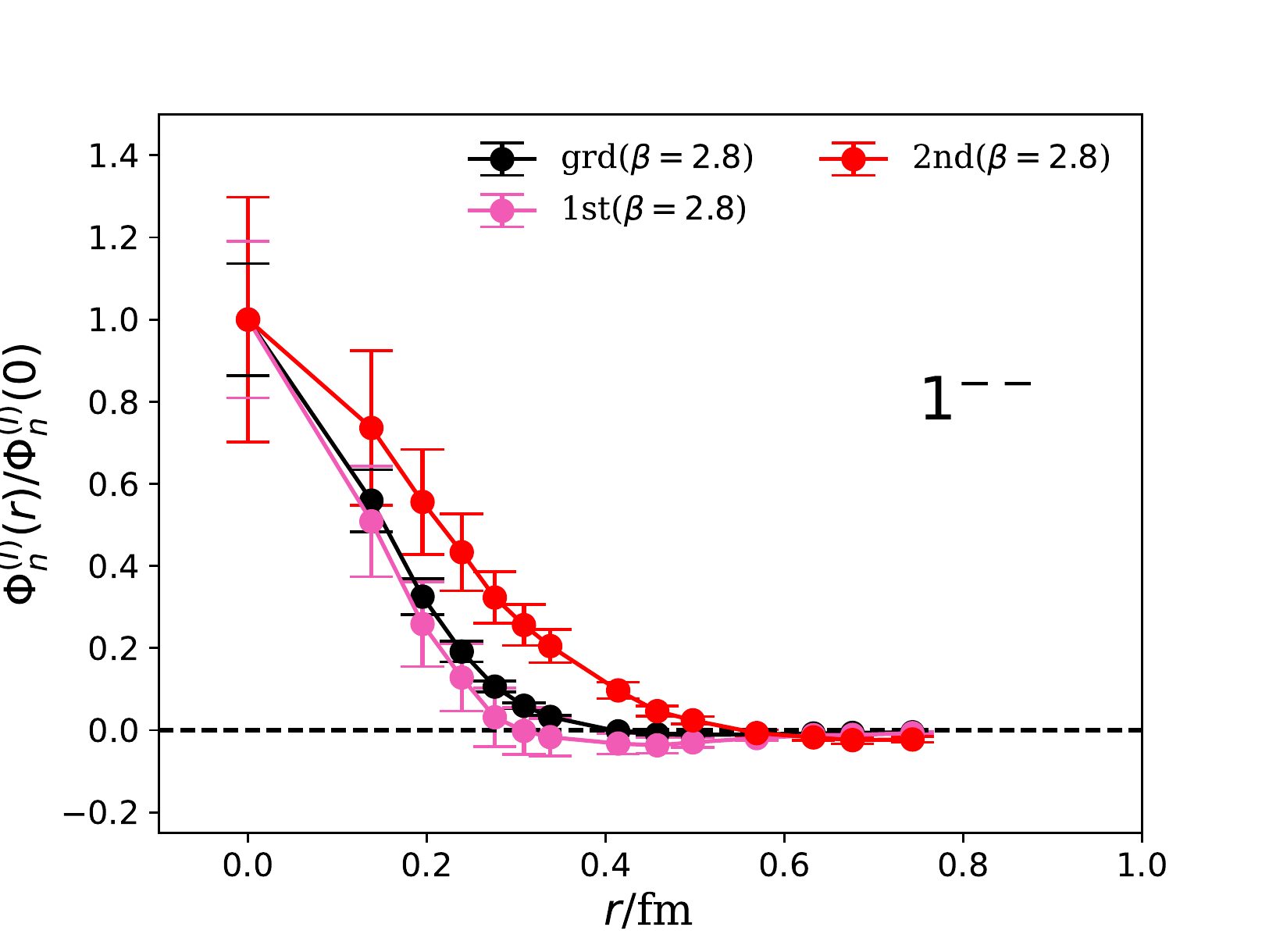}
	\caption{\label{fig:wav-PV-24} BS wave functions $\Phi_n^{(I)}(r)$'s for $0^{-+}$ (left panels) and $1^{--}$ states (right panels), which are derived from the three-mass-term fits. The upper two panels are the results at $t_\mathrm{min}/a_t=10$ for $\beta=2.4$, and the lower ones are the results at $t_\mathrm{min}/a_t=16$ for $\beta=2.8$.}
\end{figure*}
The situation for $0^{-+}$ and $1^{--}$ channels is more complicated. The ground states in the two channels are $\eta_c$ and $J/\psi$ with the $\bar{c}c$ pair being in the spin singlet and spin triplet, respectively. In contrast, recalling that the $\bar{c}c$ component of the hybrid operators defined in Eq.~(\ref{operator}) for $0^{-+}$ and $1^{--}$ are in spin triplet and spin singlet, respectively, we expect that the couplings of these operators to conventional $\bar{c}c$ states will be suppressed to an extent owing to the spin-flipping of charm quarks. Despite this kind of suppression, we observe that the lowest lying conventional $\bar{c}c$ states contribute significantly to the two-point functions in Eq.~(\ref{two-point}). We perform three-mass-term and four-mass-term fits to the correlation functions of $0^{-+}$ and $1^{--}$ channels and address the fit results in the following.

For the three-mass-term fits in both channels, we fix the upper bound $t_\mathrm{max}$ of the time window $[t_\mathrm{min},t_\mathrm{max}]$ and vary the $t_\mathrm{min}$ to monitor the stability of the fit. For $\beta=2.4$, $t_\mathrm{max}$ is set to $t_\mathrm{max}/a_t=24$, and $t_\mathrm{min}/a_t$ varies from 9 to 13.  For $\beta=2.8$ with $t_\mathrm{max}$ being fixed at $t_\mathrm{max}/a_t=39$, and $t_\mathrm{min}/a_t$ varying from 15 to 19. The fitted masses $m_n$ and $\Phi_n^{(I)}(0)$ ($n=1,2,3$) are listed in Table~\ref{tab:P-V-3-24}, where the minimal-$\chi^2$ per degree of freedom ($\chi^2/\mathrm{dof}$) values are also given for all the $t_\mathrm{min}$. The masses are converted into values with physical units using the $a_s$ in Table~\ref{tab:lattice}. We observe that the fits are all acceptable and stable in these fit ranges with reasonable $\chi^2/\mathrm{dof}$ values. The masses of the lowest two states, i.e., $m_1$ and $m_2$, are in good agreement with those of $1S$ charmonia ($\eta_c$ and $J/\psi$) and $2S$ charmonia ($\eta_c(2S)$ and $\psi(2S)$), respectively, while $m_3$ is close to the ground state masses in $1^{-+}$ and $2^{-+}$ channels. In contrast, the spectral weight magnitude of the third state, namely $\Phi_3^{(I)}(0)$, is seemingly larger than those of the lowest two states. If the third state is dominated by the charmonium-like hybrid, this type of differences in the spectral weight magnitudes complies with our previous expectation that the couplings of hybrid-like operators to conventional charmonia are suppressed due to the spin-flipping effect. Here, we provide a qualitative interpretation of the negative sign of $\Phi_2^{(I)}(0)$. As introduced in Sec.~\ref{sec:fromalism}, the correlation functions $C_R(r,t)$ are calculated using the wall-source operator in Eq.~(\ref{wall-src}), which can be viewed as $\sum\limits_{\mathbf{X},\mathbf{s}} \bar{c}(\mathbf{X})\Gamma_R c'(\mathbf{X}+\mathbf{s})$ with $c'$ being a "dressed" charm quark field $\mathbf{B}c$, such that its matrix between the vacuum and a charmonium(-like) state $H$ in its rest frame can be qualitatively expressed as
\begin{equation}
\langle 0|O^{(W)}|H\rangle \approx V_3 \int 4\pi s^2 dr \phi_H(s),
\end{equation}
where $V_3$ is the spatial volume and $\phi_H(s)$ is the BS wave function of $H$ with respect to the distance $s$ between the charm and anti-charm quark. If $H$ is a $2S$ charmonium state, we expect that $\phi_H(s)$ has a radial node (this is actually observed;, see below) such that the integral in the above equation will result in a negative sign (note that the integrand is enlarged by the factor $s^2$ for a large $s$).

Figure~\ref{fig:wav-PV-24} shows the $\Phi_n^{(I)}(r) (n=1,2,3)$ values obtained at $t_\mathrm{min}/a_t = 10$ for $\beta=2.4$ and at $t_\mathrm{min}/a_t=16$ for $\beta=2.8$, respectively. The BS wave functions are normalized as $\Phi_n^{(I)}(0)=1$. In the figure, the left-hand panels are for $0^{-+}$ states and the right ones are for $1^{--}$ states. The figures have two features: First, the $r$-dependences of $\Phi_n^{(I)}$ are similar for $0^{-+}$ and $1^{--}$ states; second, $\Phi_1^{(I)}(r)$ and $\Phi_2^{(I)}(r)$ behave almost the same with respect to $r$ and are clearly different from $\Phi_3(r)$. Since the lowest two states correspond to the $1S$ and $2S$ charmonia, the similarity of $\Phi_1^{(I)}(r)$ and $\Phi_2^{(I)}(r)$ implies that the distance $r$ between the $c\bar{c}$ component and the gluonic component is not a characteristic dynamical variable for conventional charmonia. This is in contrast to the situation of $1^{-+}$ states in that the first excited state has one radial node. These observations can be understood as follows: According to the parameterization of $C_R(r,t)$ in Eq.~(\ref{model}), the contribution of higher states almost vanish in large time regions; thus, the states appearing at large $t$ values are most likely the $1S$ and $2S$ charmonia that are the lowest states in the $0^{-+}$ and $1^{--}$ channels.

\begin{table*}[t]
	\caption{\label{tab:P-V-4-24}
		Fitted masses $m_n,n=1,2,3,4$ of $0^{-+}$ and $1^{--}$ states at different $t_\mathrm{min}$ on the two lattices, where the mass values are converted into  physical units using the lattice spacings listed in Tab.~\ref{tab:lattice}. The $\chi^2/\mathrm{dof}$ value of each fit is provided to indicate the fitting quality. The $\Phi_n^{(I)}(r)$ values at $r=0$ are also listed.}
	\begin{ruledtabular}
		\begin{tabular}{ccccclccccc}
			&$t_{\rm min}/a_t$ & $\chi^2/{\rm dof}$ & $m_1$ (GeV) & $m_2$ (GeV) & $m_3$ (GeV) &$m_4$ (GeV) & $\Phi_1^{(I)}(0)$  & $\Phi_2^{(I)}(0)$ & $\Phi_3^{(I)}(0)$ & $\Phi_4^{(I)}(0)$\\\hline
			&& & && $\beta=2.4$ & & & & &\\\hline
			$0^{-+}$	
			&	4   &  1.11     &  2.953(5)   &  4.41(8)  &  4.97(21)   & 8.4(1.3)
			&  5.89(11)   &  0.9(6.8) &  33(4)      & 24(14) \\
			&	3   &  1.13     &  2.950(3)   &  4.37(5)  &  5.07(11)   & 7.7(0.4)
			&  5.83(7)    &  3.6(2.6) &  31(1)      & 13(2)\\
			&	2   &  1.45     &  2.944(2)   &  4.39(2)  &  5.31(5)    & 8.0(0.2)
			&  5.69(4)    &  8.3(8)   &  29.6(5)    & 9.2(7) \\
			&	1   &  1.98     &  2.939(2)   &  4.45(1)  &  5.52(3)    & 8.2(0.1)
			&  5.60(3)    &  12.1(4)  & 28.0(3)     & 6.8(3)\\
			&&&&&&&&&&\\
			$1^{--}$
			&	4   &  1.23     &  3.054(6)   &  4.40(8)  &  5.04(23)   & 7.8(0.9)
			&  5.66(15)   &  5.4(5.2) &  28(3)      & 19(8) \\
			&	3   &  1.23     &  3.051(5)   &  4.33(6)  &  5.08(13)   & 7.2(0.3)
			&  5.59(11)   &  5.6(2.6) &  28(1)      & 13(2)\\
			&	2   &  1.46     &  3.046(3)   &  4.39(2)  &  5.35(5)    & 7.8(0.2)
			&  5.48(06)   &  10.2(8)  &  27.4(6)    & 9.2(8) \\
			&	1   &  1.73     &  3.040(2)   &  4.45(1)  &  5.59(3)    & 8.1(0.1)
			&  5.37(04)    &  14.1(4)  & 26.2(3)     & 6.6(3)\\\hline
				&& & && $\beta=2.8$ & & & & &\\\hline
				$0^{-+}$	
			&    6   & 1.06     &  2.981(4)  &  4.26(7)    &  5.35(11) & 8.8(0.8)
			&  12.9(2)   &  -3.8(2.4)  &  63(3)    & 47(16) \\
			&    5  &  0.77     &  2.988(4)  &  4.30(7)    &  5.23(9)  & 8.5(0.4)
			&  13.2(2)   &  -7.9(3.0)  &  63(3)    & 50(5)  \\
			&    4  &  0.79     &  2.976(4)  &  4.39(6)    &  5.40(7)  & 8.3(0.3)
			&  12.6(2)   &  -1.9(2.3)  &  61(2)    & 40(2)  \\
			&    3  &  0.84     &  2.971(3)  &  4.62(3)    &  5.58(6)  & 8.9(0.1)
			&  12.4(1)   &   3.4(2.1)  &  61(1)    & 39(2)  \\
			&    2  &  1.28     &  2.958(3)  &  4.53(2)    &  5.84(4)  & 9.0(0.1)
			&  11.9(1)   & 9.7(0.9)    & 62(1)     & 30(1)  \\
			&&&&&&&&&&\\
			$1^{--}$
			&    6   & 0.90     &  3.074(5)  &  4.28(7)    &  5.40(13) & 8.7(0.7)
			&  11.6(2)   &  -0.1(2.6)  &  61(3)    & 46(14) \\
			&    5  &  0.79     &  3.080(5)  &  4.35(7)    &  5.32(9)  & 8.7(0.4)
			&  11.9(2)   &  -2.7(3.0)  &  61(2)    & 52(6)  \\
			&    4  &  0.75     &  3.067(6)  &  4.43(6)    &  5.45(8)  & 8.2(0.2)
			&  11.3(2)   &   3.1(2.5)  &  57(2)    & 41(3)  \\
			&    3  &  0.76     &  3.064(4)  &  4.60(4)    &  5.64(6)  & 8.7(0.1)
			&  11.2(1)   &   8.8(1.7)  &  56(1)    & 39(2)  \\
			&    2  &  1.27     &  3.039(5)  &  4.52(2)    &  6.03(4)  & 9.0(0.1)
			&  10.4(1)   & 16.1(0.8)   & 59(1)     & 27(1)
		\end{tabular}
	\end{ruledtabular}
\end{table*}
\begin{figure*}[t!]
	\includegraphics[height=6cm]{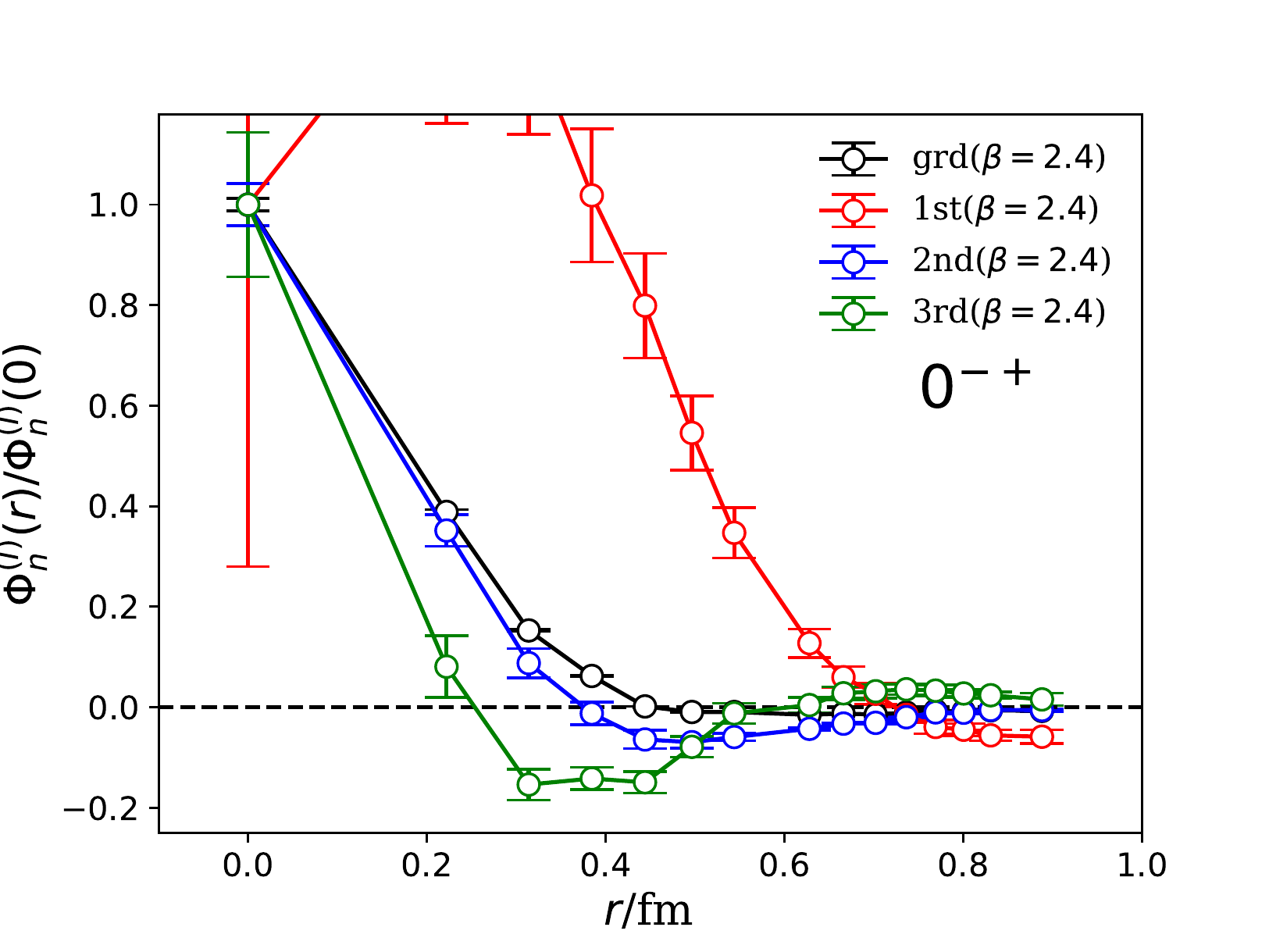}
	\includegraphics[height=6cm]{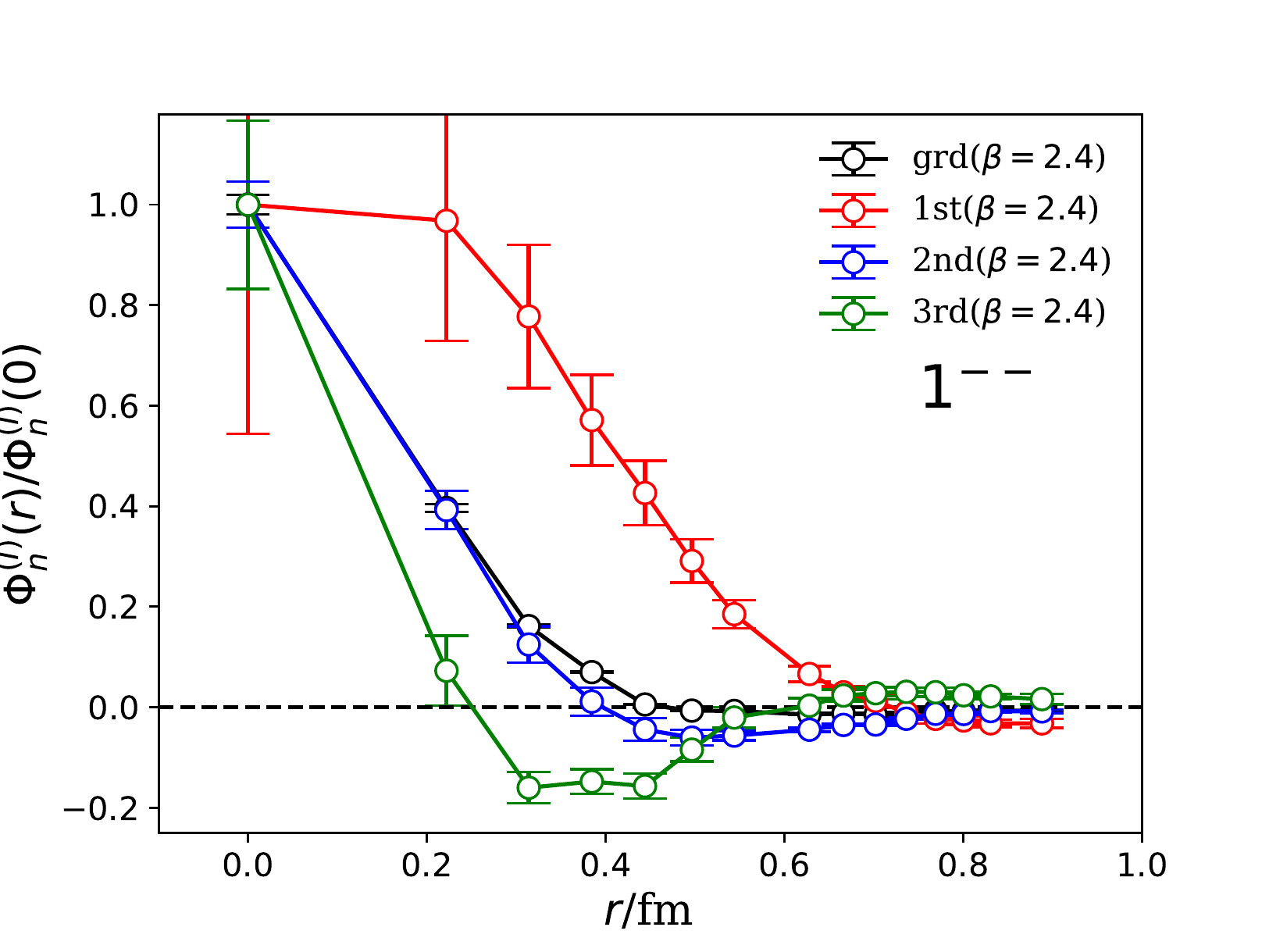}
	\includegraphics[height=6cm]{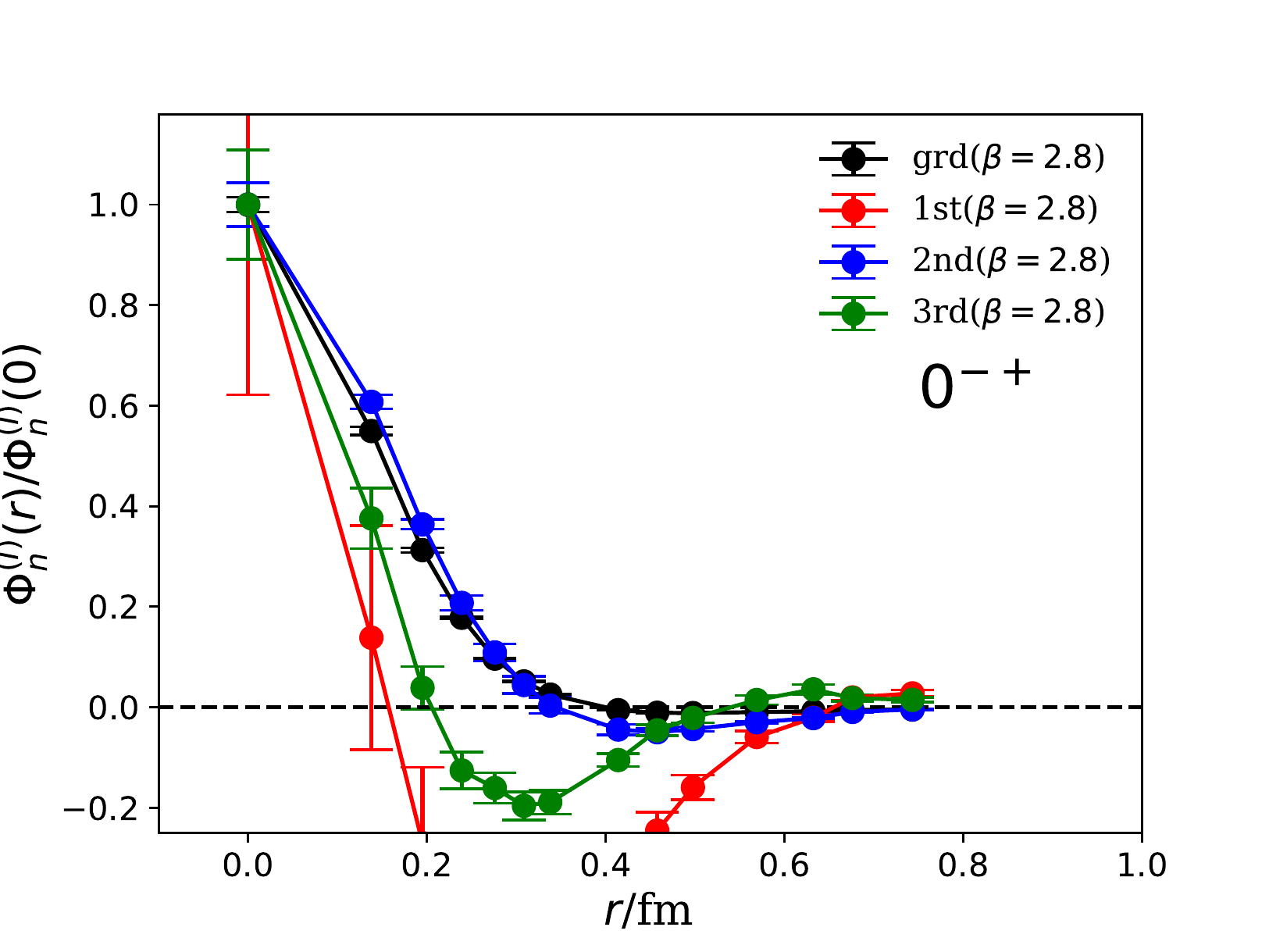}
	\includegraphics[height=6cm]{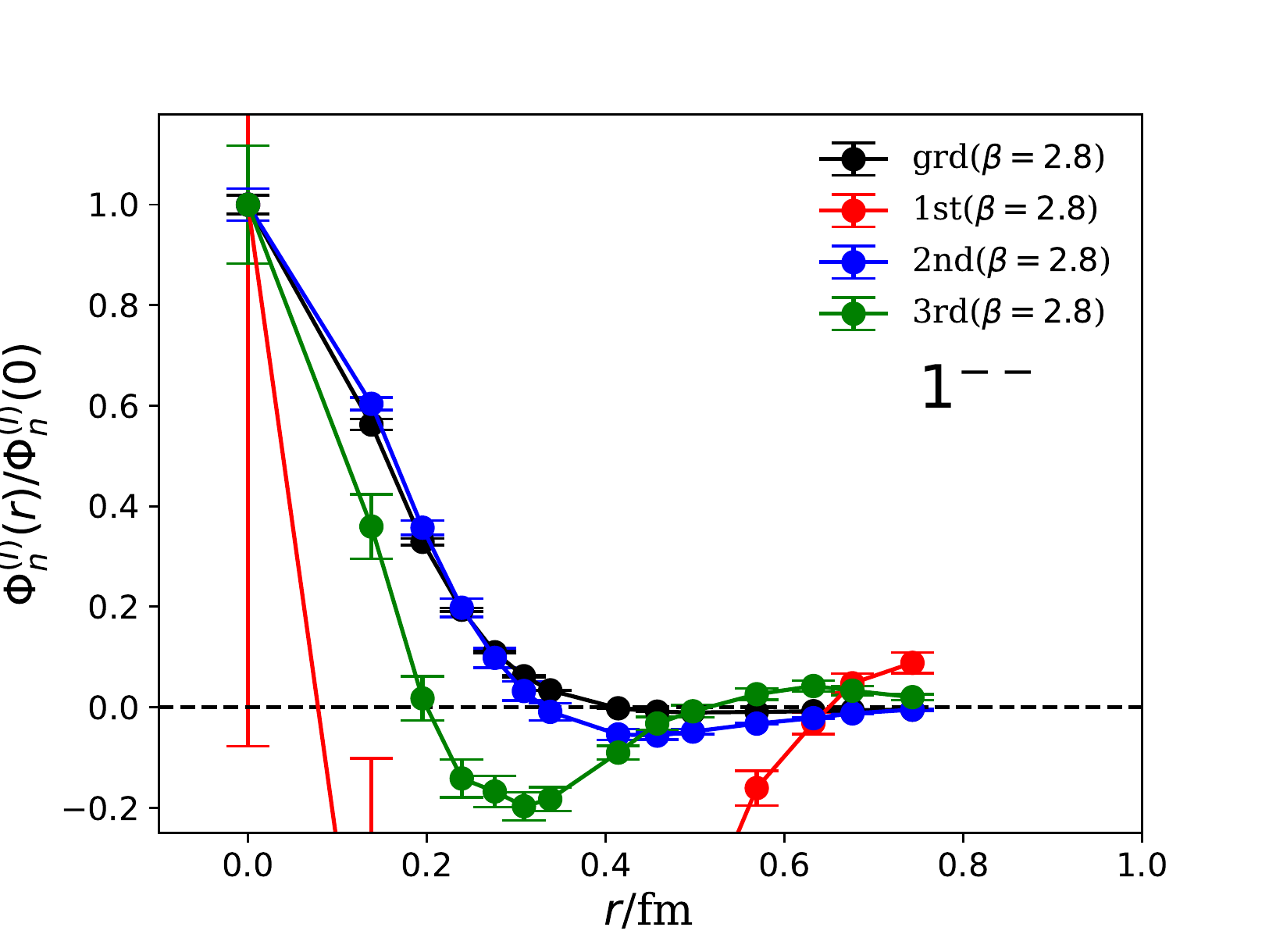}
	\caption{\label{fig:P-V-wav4} BS $\Phi_n^{(I)}(r)$'s for $0^{-+}$ functions (left-hand panels) and $1^{--}$ states (right-hand panels), which are derived from the four-mass-term fits. The upper two panels are the results at $t_\mathrm{min}/a_t=3$ for $\beta=2.4$, and the lower ones are the results at $t_\mathrm{min}/a_t=5$ for $\beta=2.8$. $\Phi_2^{(I)}(r)$ is unstable for different $t_\mathrm{min}$'s and has very large errors due to the mixing among the conventional and hybrid states nearby. The $r$-behaviors of $\Phi_3^{(I)}(r)$ and $\Phi_4^{(I)}(r)$ are stable for different $t_\mathrm{min}$'s and are similar to those of $\Phi^{(I)}_2(r)$ and $\Phi^{(I)}_3(r)$ of $(1,2)^{-+}$ states (see Fig.~\ref{fig:1-+res} and Fig.~\ref{fig:2-+res}).  }
\end{figure*}

We now discuss the results of the four-mass-term fits. To incorporate the data points in the small $t$ region, we add the fourth mass term in the fit model. Table~\ref{tab:P-V-4-24} shows the fit results. We observe that the four-mass-term model closely fits the data for $t_\mathrm{min}/a_t$ starting even from $t_\mathrm{min}/a_t=1$ for $\beta=2.4$ and $t_\mathrm{min}/a_t=2$ for $\beta=2.8$. The ground state masses are compatible with the $1S$ charmonia ($\eta_c$ for $0^{-+}$ and $J/\psi$ for $1^{--}$) but are slightly lower than the results of the three-mass-term fits. The masses of the second lowest states ($m_2$) are approximately $4.4$ GeV and significantly higher than the expected $2S$ charmonium masses. The masses of the third and the fourth states ($m_3$ and $m_4$) are close to those of the second and third states
in the $1^{-+}$ and $2^{-+}$ channels (see Table~\ref{tab:1-+mass} and \ref{tab:2-+mass}). For the BS wave functions at the origin $r=0$, $\Phi_1^{(I)}(0)$ is very stable with respect to the varying of $t_\mathrm{min}/a_t$ but slightly smaller than the three-mass-term fit results (this may be due to the smaller fitted masses $m_1$) whereas $\Phi_2^{(I)}(0)$ changes dramatically. Because there are a few conventional charmonium states at approximately 4 GeV, e.g., $\psi(3686)$, $\psi(4040)$, $\psi(4415)$ etc. and a would-be charmonium-like hybrid state in the $1^{--}$ channel, the second state may be an admixture of these states, such that its fitted spectral weight $\Phi_2(r)$ is very sensitive to the fit window. In contrast to $\Phi_2^{(I)}(0)$, $\Phi_3^{(I)}(0)$ and $\Phi_4^{(I)}(0)$ are stable for different fits
and are larger than $\Phi_1^{(I)}(0)$ which implies that the third and fourth states contribute significantly to the correlation function $C_R(r,t)$ in the small $t$ region despite their high masses. The BS wave functions obtained from four-mass-term fits are plotted in Fig.~\ref{fig:P-V-wav4}, where the plots in the first row are results for $t_\mathrm{min}/a_t=3$ for $\beta=2.4$ and the plots in the second row are those for $t_\mathrm{min}/a_t=5$. First, the $\Phi_n(r)$ values of $0^{-+}$ and $1^{--}$ states are similar to each other. Second, the $\Phi_n^{(I)}(r)$ values of the two lattices ($\beta=2.4$ and $\beta=2.8$) exhibit a slight finite $a_s$ dependence. The $r$-behavior of $\Phi_1^{(I)}(r)$ is very similar to that of the three-mass-term fits. As indicated above, the second state may be an admixture of multiple states of conventional $nS$ charmonia and the possible hybrid state; we do not consider the $\Phi_2^{(I)}(r)$ values to be significant. $\Phi_3^{(I)}(r)$ and $\Phi_4^{(I)}(r)$ have similar features to $\Phi_2^{(I)}(r)$ and $\Phi_3^{(I)}(r)$ in $1^{-+}$ and $2^{-+}$ channels.

\subsection{Joint discussion of $1^{--}$ and $(0,1,2)^{-+}$ states}
Until now, we have provided a detailed description of the fitting procedures and fitted results of the correlation functions involving the $c\bar{c}-g$ operators in the $1^{--}$ and $(0,1,2)^{-+}$ channels.
The results of the $1^{-+}$ and $2^{-+}$ are solid and have slight ambiguities. The situation for $0^{-+}$ and $1^{--}$ channels is a little more complicated since both conventional charmonia and possible charmonium-like hybrids contribute to correlation functions. In the three-mass-term fits for both channels, at larger $t_{\rm min}$ values of the fit window, the ground states are stable with masses consistent with those of $J/\psi$ and $\eta_c$, the second states have masses of approximately $3.7$ GeV, while the third states have masses of roughly $4.5$ GeV with larger spectral weights. We consider the lowest two states to be the $1S$ and $2S$ conventional charmonia, and tentatively assign the third state in each channel to be the corresponding ground state charmonium-like hybrid. In the four-mass-term fits with smaller $t_{\rm min}$ values, for each channel, while the lowest state does not change much, the mass of the second state varies slightly, and the corresponding wave function changes drastically. This can be attributed to the interplay of conventional $nS$ charmonia and the ground state hybrid. The third state is relatively stable with varying $t_{\rm min}$ and has a mass close to those of the second states in $(1,2)^{-+}$ channels; therefore, we consider it to be the first excited hybrid state in the corresponding channel. The $r$-dependence of the BS wave functions reinforces the above assignments: the wave functions of the three states in the three-mass-term fit have no nodes in the $r$-direction, and those of the third and fourth states in the four-mass-term fit have one and two nodes, respectively.

\begin{figure}[t!]
	\includegraphics[height=6cm]{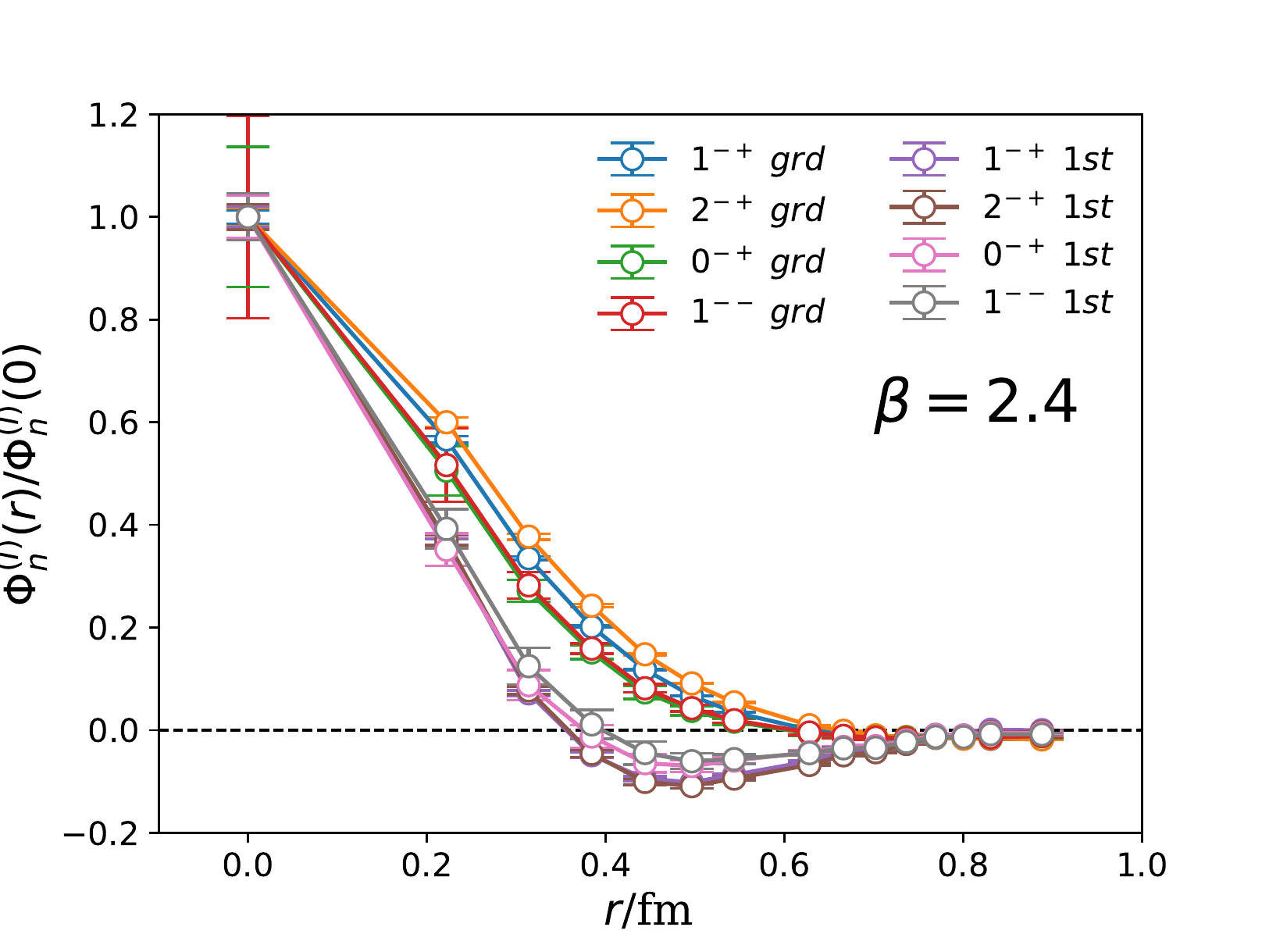}
	\includegraphics[height=6cm]{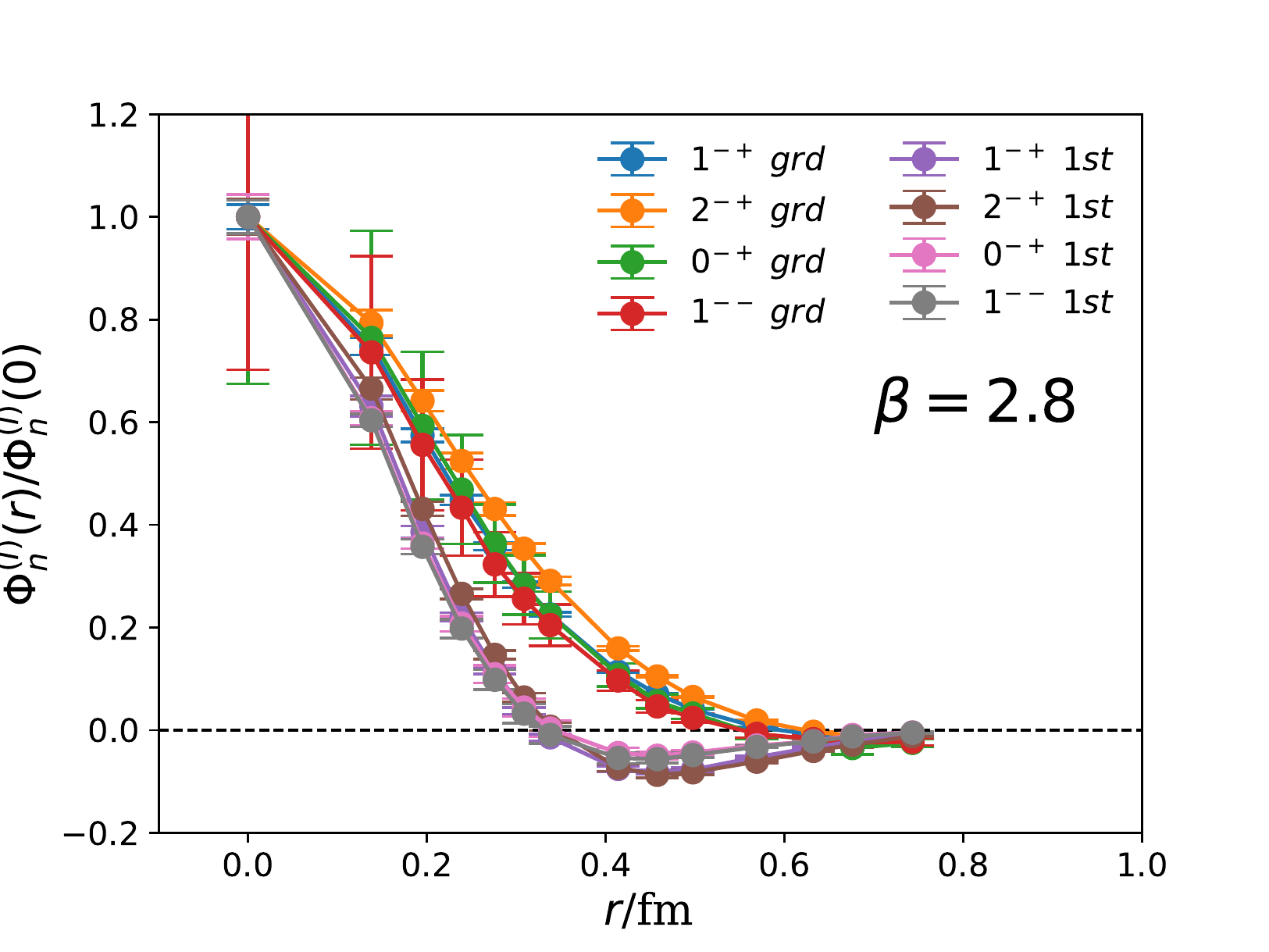}
	\caption{\label{wave-all} BS wave functions (normalized by $\Phi_n^{(I)}(0)$) of the $1^{--}$
	and $(0,1,2)^{-+}$ multiplets with masses of approximately 4.3-4.5 GeV (0 node) and 5.4-5.7 GeV
	(one node). The upper and lower panels are for $\beta=2.4$ and $\beta=2.8$ lattices,
	respectively. We observe that the BS wave functions of the states in each
	multiplet almost fall onto each other.}
\end{figure}

The fitted mass spectrum is shown in Table~\ref{spect}, in which the values are averages of the two lattices. After singling out the states that correspond to the conventional $\bar{c}c$ states in $0^{-+}$ and $1^{--}$ channels, the masses of the states whose BS wave functions have the same number of nodes are listed in the same row for all the four channels. We observe that the masses of the 0-node and 1-node states are nearly degenerate (we do not consider the masses of the 2-node states to be significant since they may have substantial contamination from
higher states). The BS wave functions of the 0-node states (except for the states corresponding the
conventional charmonia in $0^{-+}$ and $1^{--}$ channels) and 1-node states in the four
channels are plotted in Fig.~\ref{wave-all} (the upper panel for $\beta=2.4$ and the lower one
for $\beta=2.8$). The striking observation is that the wave functions of different channels
almost fall onto each other.

 \begin{table}[t]
 	\centering \caption{\label{spect}
 	Mass spectrum of the $1^{--}$ and $(0,1,2)^{-+}$ states. The lowest three masses in the $1^{--}$ and $0^{-+}$ channels are from the three-mass-term fits. The other two masses in the two channels are the masses of the third and fourth states, respectively, from the four-mass-term fits. Except for masses of the
 	lowest two states in $1^{--}$ and $0^{-+}$ channels which correspond to conventional
 	charmonia, other masses are arranged in each row by the number of nodes (\#node) of the
 	BS wave functions of related states. The values are averages of the two lattices and are converted into the physical units using the lattice spacings listed in
 	 Tab.~\ref{tab:lattice}.}
 	\begin{ruledtabular}
 		\begin{tabular}{cllll}
 			\#node  & $m(1^{--})$ & $m(0^{-+})$ & $m(1^{-+})$  & $m(2^{-+})$  \\
 			& (GeV)       & (GeV)       & (GeV)        &  (GeV)    \\
 			\hline
 			0   &  3.109(5)    & 3.010(4)    &    \--     &    \--    \\
 			0   &  3.703(82)   & 3.672(76)   &    \--     &    \--    \\
 			&&&&\\
 			0   &  4.591(69)   & 4.551(63)   &  4.309(2)  &  4.419(3) \\
 			1   &  5.460(31)   & 5.393(28)   &  5.693(12) &  5.779(12)\\
 			2   &  8.226(99)   & 8.286(109)  &  7.661(31) &  7.708(29)
 		\end{tabular}
 	\end{ruledtabular}
 \end{table}

\section{Results from type-II operators }
As a self-consistent check, another type of BS wave function $\Phi^{(II)}_n(r)$ of the $1^{--}, (0,1,2)^{-+}$ charmonium-like states is explored by implementing the type-II operators $O_R^{(II)}(r,t)$ defined in Eq.~(\ref{eq:type-II}), where $r$ is the spatial separation between the charm quark field $\bar{c}$ and $c\mathbf{B}$ component. The corresponding correlation functions are
expressed as
\begin{eqnarray}\label{model-II}
C_R(r,t)&\equiv& \frac{1}{N_\tau}\sum\limits_{\tau}\langle O_R^{(II)}(r,t+\tau)O_R^{(W)\dagger}(\tau)\rangle\nonumber\\
&=& \sum\limits_n \Phi^{(II)}_n(r)e^{-m_n t},
\end{eqnarray}
The data fitting strategy is the same as that for type-I operator, and will not be repeated here. We directly present the results (this part of the calculation was carried out only on the $\beta=2.4$ lattice).
\begin{table*}[t]
 	\caption{\label{tab:spec_1-II} Fitted masses of $0^{-+}$ states at different $t_\mathrm{min}$ values through the two-mass-term (top part), the three-mass-term ( middle part) and the four-mass-term fits to the correlation functions involving type-II operators. The mass values are converted into physical units using the lattice spacings listed in Tab.~\ref{tab:lattice}. The $\chi^2/\mathrm{dof}$ of each fit is provided to indicate the fitting quality. $\Phi_n^{(II)}(r)$ values at $r=0$ are also listed.
 		}
 	\begin{ruledtabular}
 		\begin{tabular}{cccccccccc}
 			$t_{\rm min}/a_t$ & $\chi^2/{\rm dof}$ & $m_1$ (GeV) & $m_2$ (GeV) & $m_3$ (GeV) & $m_4$ (GeV)
 			&$\Phi_1^{(II)}(0)$      & $\Phi_2^{(II)}(0) $       & $\Phi_3^{(II)}(0)$ &$\Phi_4^{(II)}(0)$ \\\hline
 			20   &  0.51     &  2.988(1)  &  3.709(37) &  \--  &\--&  14.0(0.1)  &  -10.0(1.0) &\--  & \--  \\
 			19   &  0.55     &  2.987(1)  &  3.703(33) &  \--  &\--&  13.9(0.1)  &  -8.3(0.8)  &\--  & \--  \\
 			18   &  0.58     &  2.987(1)  &  3.667(27) &  \--  &\--&  13.9(0.1)  &  -6.8(0.5)  &\--  & \--  \\
 			17   &  0.59     &  2.987(1)  &  3.655(23) &  \--  &\--&  13.9(0.1)  &  -6.4(0.4)  &\--  & \--  \\
 			16   &  0.56     &  2.985(1)  &  3.687(19) &  \--  &\--&  13.7(0.1)  &  -5.6(0.3)  &\--  & \--  \\
 			&&&&&&&&&\\
 			11   &  0.99     &  2.988(4)  &  3.678(63) &  4.49(19)&\-- & 14.0(0.4) & -10(2) & 50(11) &\-- \\
 			10   &  0.89     &  2.987(3)  &  3.714(43) &  4.58(13)&\-- & 13.9(0.3) & -10(2) & 59(8)  &\-- \\
 			9    &  0.76     &  2.986(3)  &  3.719(35) &  4.66(10)&\-- & 13.9(0.3) & -10(2) & 65(6)  &\-- \\
 			8    &  0.83     &  2.996(5)  &  3.586(41) &  4.57(5) &\-- & 14.8(0.5) & -10(1) & 58(3)  &\-- \\
 			7    &  0.98     &  2.980(3)  &  3.678(24) &  4.88(5) &\-- & 13.4(0.2) & -5(1)  & 73(2)  &\-- \\	
 			&&&&&&&&&\\
 			4    &  1.06     &  2.926(7)  &  3.956(49) &  5.43(47)& 5.93(48)&10.5(3)&10(3)&94(112)&-24(119)\\
 			3    &  0.99     &  2.936(7)  &  3.924(38) &  5.03(11)& 6.44(9) &11.0(3)&5(3) &58(3)  &24(6)   \\
 			2    &  2.27     &  2.928(5)  &  3.966(20) &  5.15(6) & 6.88(5) &10.7(2)&8(2) &61(1)  &23(2)   \\
 		\end{tabular}
 	\end{ruledtabular}
\end{table*}
\begin{figure}
 	\includegraphics[height=6cm]{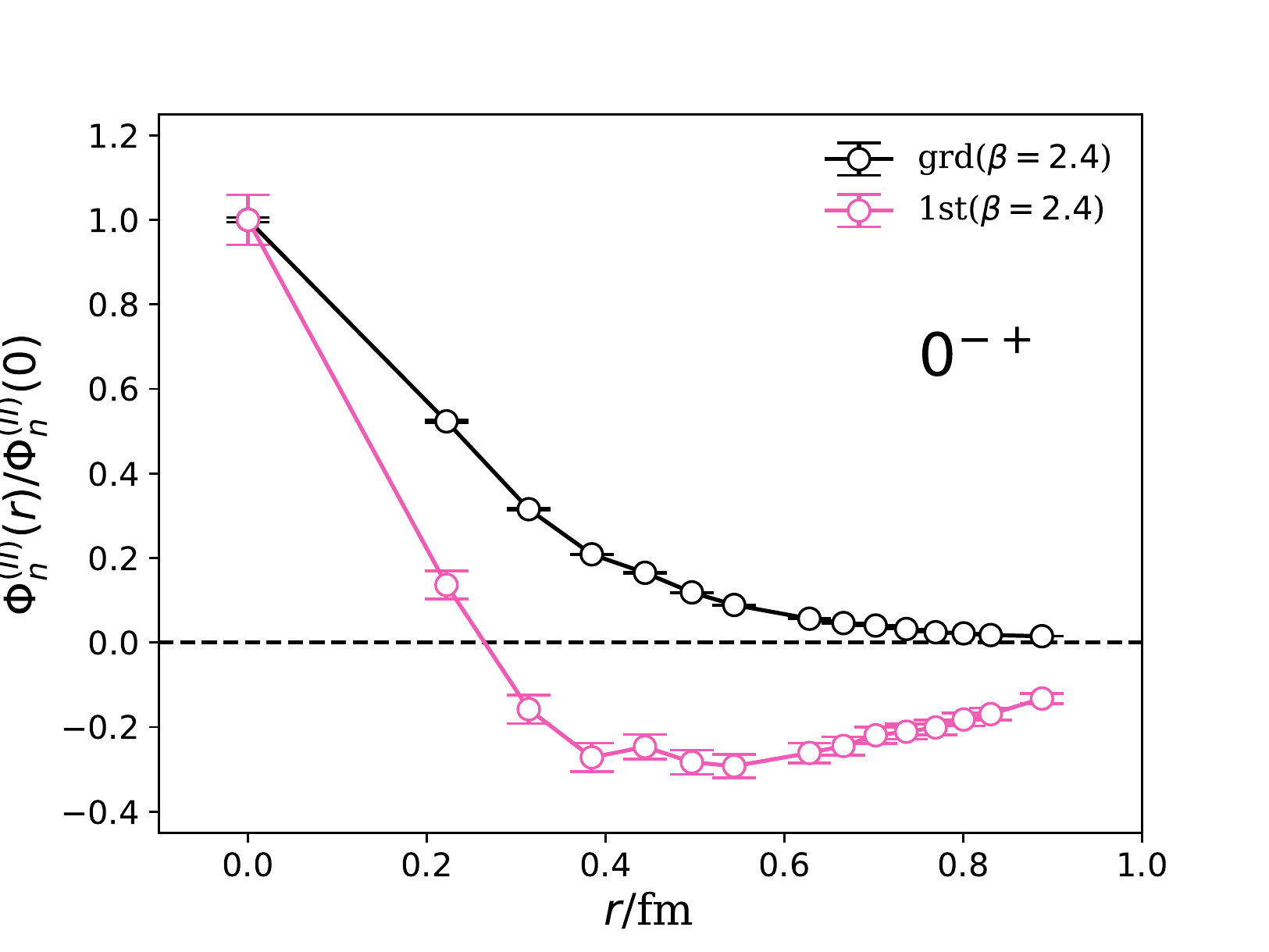}
 	\includegraphics[height=6cm]{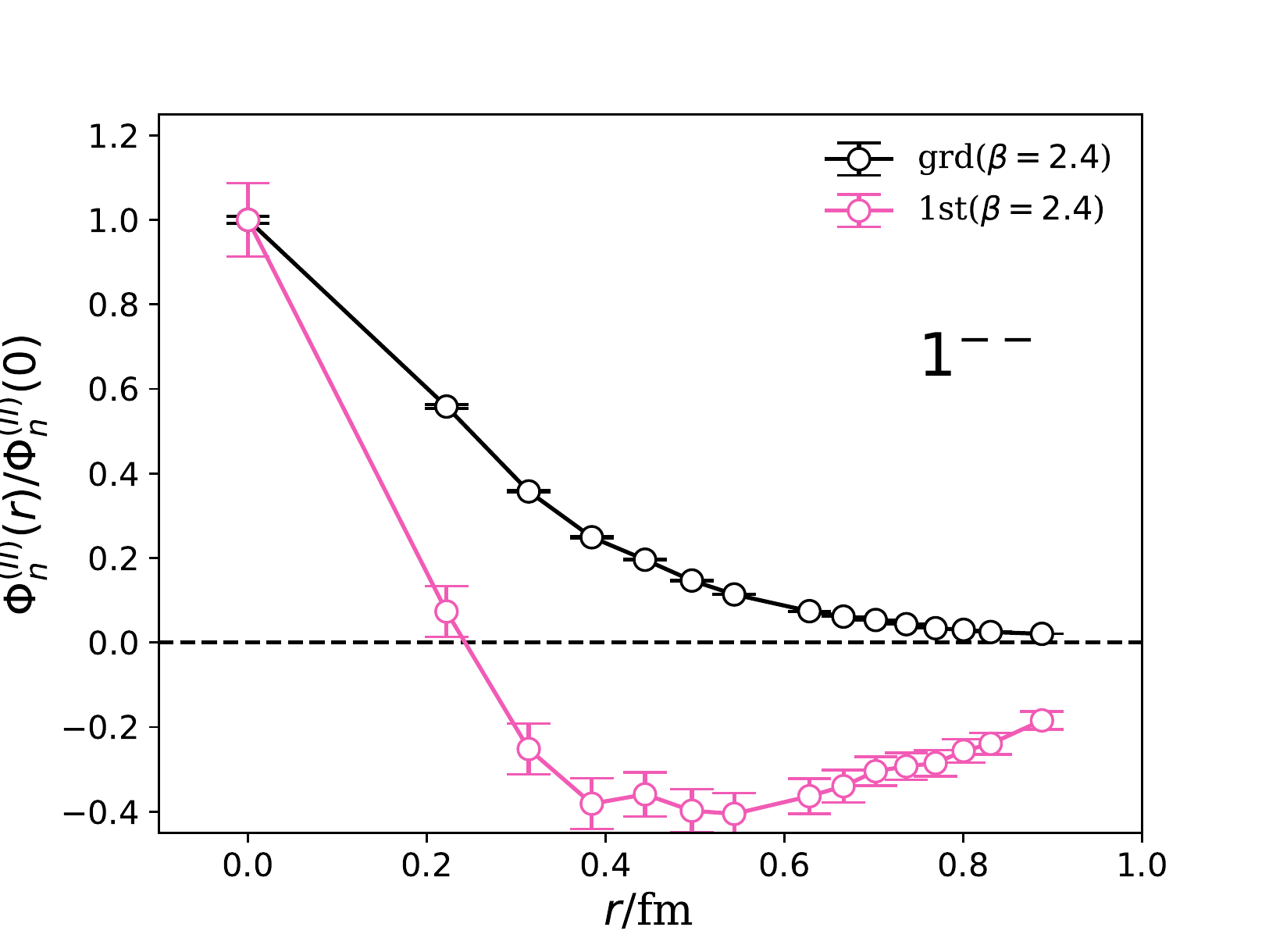}
 	\caption{Two-mass-term fit results of $\Phi_n^{(II)}(r)$ (normalized by $\Phi_n^{(II)}(0)=1$) of $0^{-+}$ (upper panel) and $1^{--}$ (lower panel) states using type-II sink operators. The $x$-axis is the separation between the $\bar{c}B$ and $c$ components. The fits are performed with $t_{\rm min}/a_t=17$ and $t_{\rm max}/a_t=39$ with the $\chi^2/{\rm dof}$ being 0.59 and 0.60 for $0^{-+}$ and $1^{--}$ channels, respectively.\label{figure6} }
\end{figure}
\begin{figure}
	\includegraphics[height=6cm]{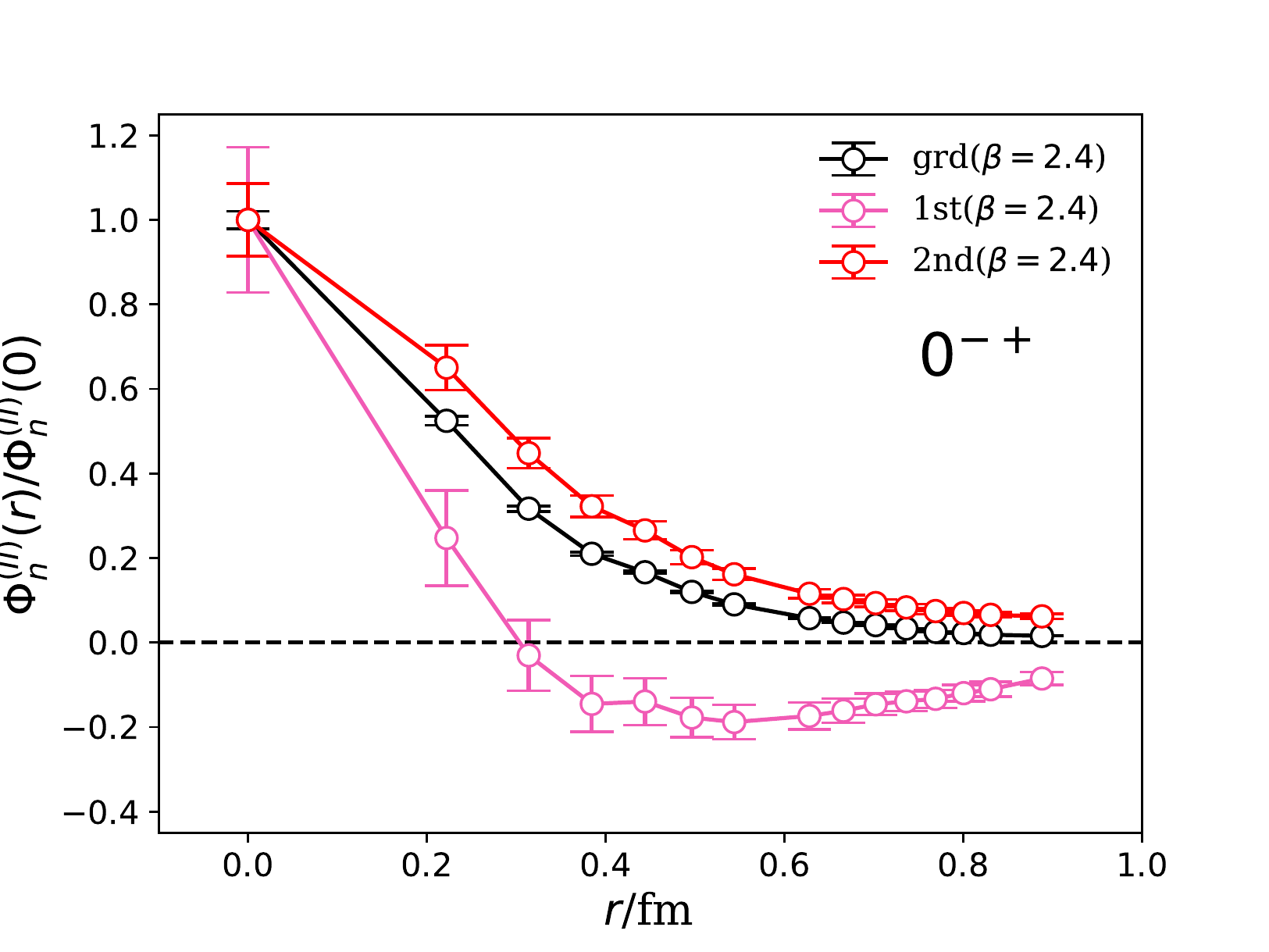}
	\includegraphics[height=6cm]{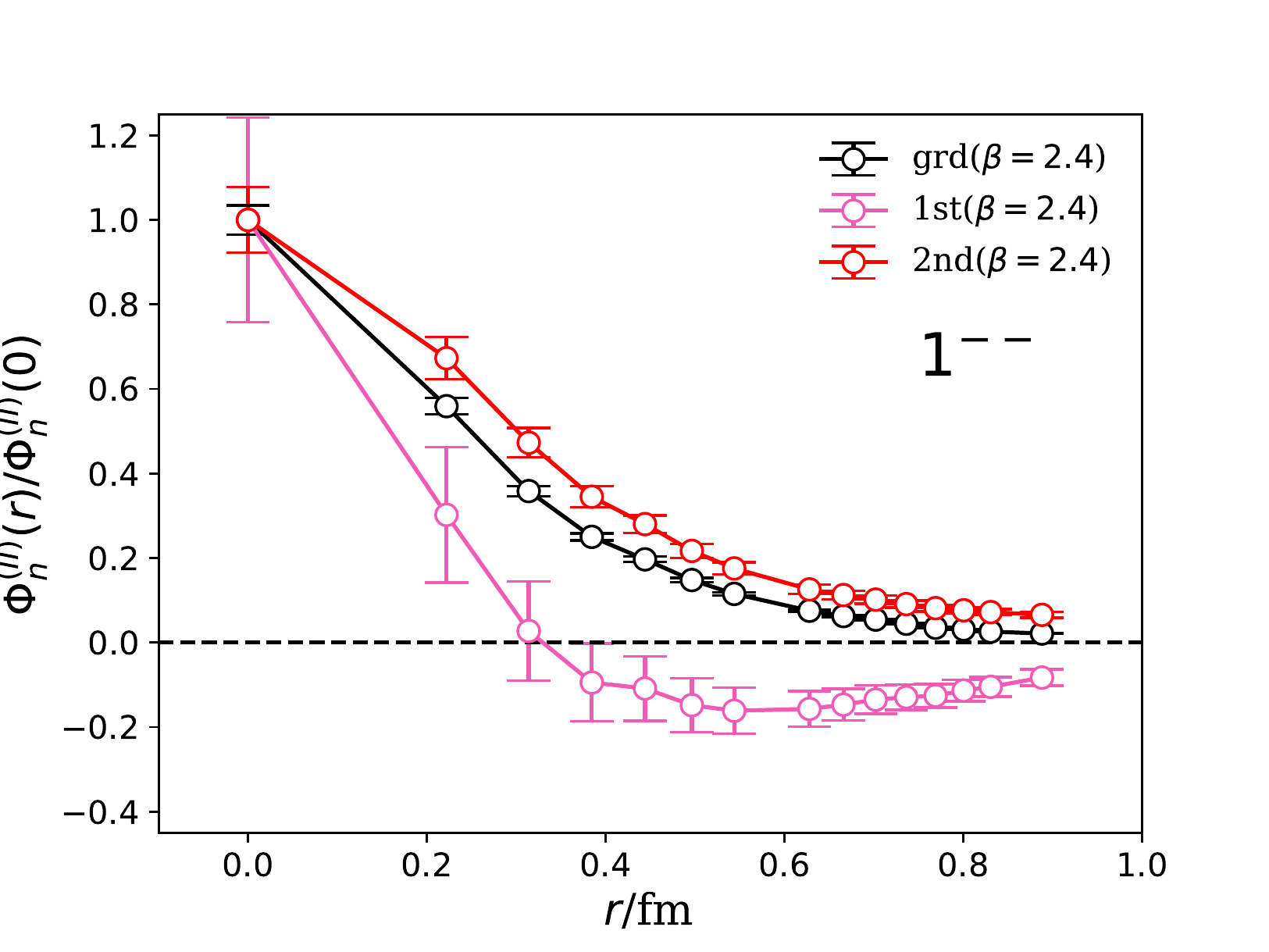}
	\caption{Three-mass-term fit results of $\Phi_n^{(II)}(r)$ (normalized by $\Phi_n^{(II)}(0)=1$) of $0^{-+}$ (upper panel) and $1^{--}$ (lower panel) states using type-II sink operators. The $x$-axis is the separation between the $\bar{c}B$ and $c$ components. The fits are performed with $t_{\rm min}/a_t=9$ and $t_{\rm max}/a_t=24$ with
		$\chi^2/{\rm dof}$ are 0.76 and 0.74 for $0^{-+}$ and $1^{--}$ channels, respectively.\label{figure7} }
\end{figure}
\begin{figure}
	\includegraphics[height=6cm]{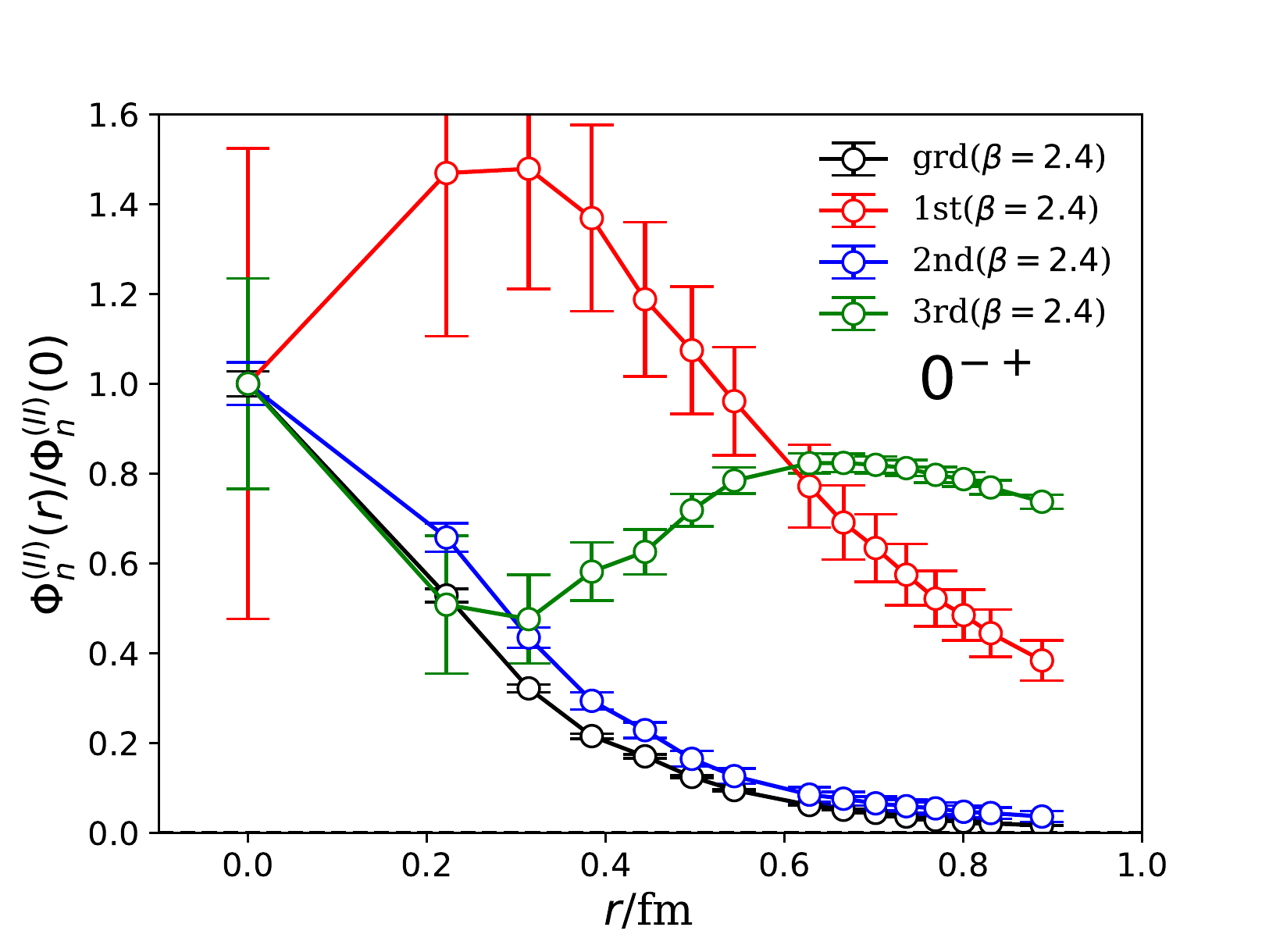}
	\includegraphics[height=6cm]{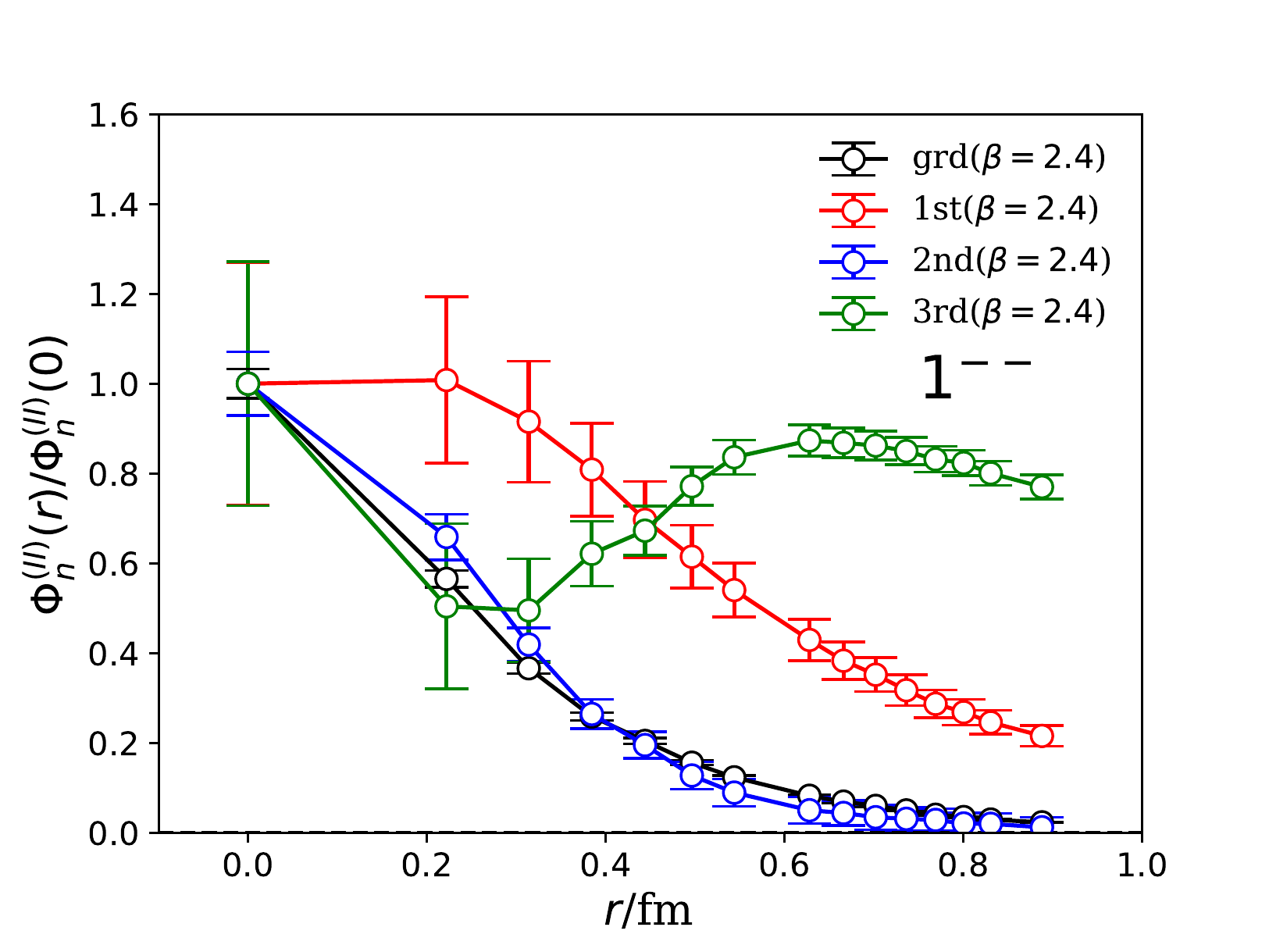}
	\caption{Four-mass-term fit results of $\Phi_n^{(II)}(r)$ (normalized by $\Phi_n^{(II)}(0)=1$) of $0^{-+}$ (upper panel) and $1^{--}$ (lower panel) states using type-II sink operators. The $x$-axis is the separation between $\bar{c}B$ and $c$ components. The fits are performed with $t_{\rm min}/a_t=3$ and $t_{\rm max}/a_t=15$ with $\chi^2/{\rm dof}$ are 0.99 and 1.04 for $0^{-+}$ and $1^{--}$ channels, respectively.\label{figure8} }
\end{figure}
\begin{table*}[t]
 	\caption{\label{tab:1mm_cgc} Fitted masses of $1^{--}$ states at different $t_\mathrm{min}$ values through the two-mass-term (top part), the three-mass-term ( middle part) and the four-mass-term fits to the correlation functions involving type-II operators. The mass values are converted into physical units using the lattice spacings listed in Tab.~\ref{tab:lattice}. The $\chi^2/\mathrm{dof}$ of each fit is provided to indicate the fitting quality. $\Phi_n^{(II)}(r)$ values at $r=0$ are also listed.
 		}
 	\begin{ruledtabular}
 		\begin{tabular}{cccccccccc}
 			$t_{\rm min}/a_t$ & $\chi^2/{\rm dof}$ & $m_1$ (GeV) & $m_2$ (GeV) & $m_3$ (GeV) & $m_4$ (GeV)
 			&$\Phi_1^{(II)}(0)$      & $\Phi_2^{(II)}(0) $       & $\Phi_3^{(II)}(0)$ &$\Phi_4^{(II)}(0)$ \\\hline
            20 &  0.53 &  3.090(1) &  3.763(49) &  \--  &\--  &  13.6(0.1)  &  -8.0(1.1)  &  \--  &  \-- \\
            19 &  0.57 &  3.090(1) &  3.717(43) &  \--  &\--  &  13.6(0.1)  &  -6.4(0.7)  &  \--  &  \-- \\
            18 &  0.60 &  3.089(1) &  3.716(36) &  \--  &\--  &  13.4(0.1)  &  -5.0(0.5)  &  \--  &  \-- \\
            17 &  0.60 &  3.088(1) &  3.704(30) &  \--  &\--  &  13.4(0.1)  &  -4.6(0.4)  &  \--  &  \-- \\
            16 &  0.56 &  3.086(1) &  3.746(26) &  \--  &\--  &  13.1(0.1)  &  -3.6(0.4)  &  \--  &  \-- \\
 			&&&&&&&&&\\
            11 &  0.86 &  3.096(8) &  3.731(92) &   4.37(20) &\--  &  14.1(0.9)  &  -13(5)  &  47(7)  &  \-- \\
            10 &  0.80 &  3.097(7) &  3.694(65) &   4.47(12) &\--  &  14.3(0.7)  &  -11(3)  &  51(5)  &  \-- \\
            9  &  0.74 &  3.092(5) &  3.733(45) &   4.65(10) &\--  &  13.7(0.5)  &  -8(2)  &  62(5)   &  \-- \\
            8  &  0.77 &  3.099(7) &  3.649(48) &   4.63(6)  &\--  &  14.4(0.7)  &  -9(2)  &  60(2)   &  \-- \\
            7  &  0.96 &  3.078(4) &  3.732(28) &   4.93(5)  &\--  &  12.6(0.3)  &  -3(1)  &  73(3)   &  \-- \\	
 			&&&&&&&&&\\
            4  &  1.05 &  3.018(9) &  4.018(54) &   5.37(40) &   5.97(41) &9.7(0.4)   &13(4)  &70(49)  &1(55) \\
            3  &  1.04 &  3.024(8) &  4.028(38) &   5.13(14) &   6.36(11) &10.0(0.3)  &11(3)  &53(4)   &24(7) \\
            2  &  1.91 &  3.020(6) &  4.049(22) &   5.22(7)  &   6.87(5)  &9.8(0.2)   &13(2)  &57(1)   &22(3) \\
 		\end{tabular}
 	\end{ruledtabular}
 \end{table*}

\subsection{$0^{-+}$ and $1^{--}$ states}
 Let us begin with the results of the $0^{-+}$ and $1^{--}$ channels. The results of $m_n$ and $\Phi_n^{(II)}$ from the two-mass-term, three-mass-term and four-mass-term fits in the $0^{-+}$ channel, for which the masses are converted into values with physical units, are presented; the $\chi^2/\mathrm{dof}$ of each fit is also presented to indicate the fit quality. The results in the $1^{--}$ channel are similar, which can be checked by comparing Table \ref{tab:spec_1-II} and Table \ref{tab:1mm_cgc}; therefore, we omit these to save space.

 When $t_\mathrm{min}/a_t>15$, the $C_R(r,t)$ functions can be fitted using two mass terms with reasonable $\chi^2/\mathrm{dof}$ values, as shown in the first five rows in Table~\ref{tab:spec_1-II}. The fitted masses are stable versus $t_\mathrm{min}$ and approximately $m_1\approx 2.987(1)$ GeV and $m_2\approx 3.654(23)$ GeV at $t_\mathrm{min}/a_t=17$, which are consistent with those of $\eta_c(1S)$ and $\eta_c(2S)$. The wave functions $\Phi_n^{(II)}(r)$ of $0^{-+}$ and $1^{--}$ states are shown in Fig.~\ref{figure6}, where $\Phi_n^{(II)}(r)$ is normalized using $\Phi_n^{(II)}(0)=1$. It is interesting to observe that $\Phi_1^{(II)}(r)$ has no node in the $r$-direction, whereas $\Phi_2^{(II)}(r)$ has one radial node. This type of $r$-behaviors is
 in qualitative agreement with the expectation of the non-relativistic quark model and can be understood as follows. First, we assume the non-relativistic approximation is reasonable for charm quark systems to a certain extent. Since operator
 $O_R^{(II)}(r,t)$ has a spatial structure in that the charm field $\bar{c}$ is spatially separated from block $c\mathbf{B}$ by distance $r$, and the BS wave function is defined by $\Phi^{(II)}_n(r)\propto \langle \Omega|O_R^{(II)}(r)|n\rangle$, $\Phi^{(II)}_n(r)$ can be interpreted as the probability amplitude of annihilating an anti-charm quark at the origin by $\bar{c}$ and annihilating a charm quark at $r$ by $c\mathbf{B}$ in the state $|n\rangle$. In other words, block $c\mathbf{B}$ can be considered to be a dressed charm quark field belong to the fundamental representation of the color $SU(3)$ group. Thus $\Phi^{(II)}_n(r)$ qualitatively reflects the non-relativistic wave functions of the corresponding $c\bar{c}$ state and thereby has the expected nodal structure.

 When $t_{\rm min}$ decreases to the range $7a_t \le t_{\rm min}\le 11a_t$, the third mass term is required in the fit function to describe the lattice data of $C_R(r,t)$. The results of the three-mass-term fits are tabulated in the middle part of Table~\ref{tab:spec_1-II}, where the $\chi^2/\mathrm{dof}$ values manifest the good quality of the fits in these time range. The mass values (in physical units) of $m_1, m_2, m_3$ almost do not change with respect to the change in $t_\mathrm{min}$ and  $\Phi_n^{(II)}(r=0)$'s are also very stable. The values of $m_1$ and $m_2$ are consistent with
 the result of the two-mass-term fits and are compatible with the masses of $\eta_c(1S)$ and $\eta_c(2S)$. $m_3$  is approximately 4.6(1) GeV, which is close to the result of the three-mass-term fits to the correlation functions of the Type-I operator (see Table~\ref{tab:P-V-3-24}). Note that the values of $\Phi_n^{(II)}(0)$
 are approximately twice as large as those of $\Phi_n^{(II)}(0)$ in Table~\ref{tab:P-V-3-24} for $0^{-+}$ states. This is expected since $O_R^{(II)}(r=0,t)=2O_R^{(I)}(r=0,t)$ according to the definitions in Eq.~(\ref{eq:type-I}) and ~(\ref{eq:type-II}). In contrast, $|\Phi_3^{(II)}(0)|$ is several times larger than $\Phi_1^{(II)}(0)$ and $\Phi_2^{(II)}(0)$, as is expected as $c\bar{c}g$ operators couple more to hybrid states than conventional $c\bar{c}$ states since the $c\bar{c}$ component in the $0^{-+}$ $c\bar{c}g$ operator is in spin-triplet. The wave functions $\Phi_n^{(II)}(r), n=1,2,3$ of $0^{-+}$ and $1^{--}$ states are shown in Fig.~\ref{figure7}, where $\Phi_n^{(II)}(r)$ is normalized using $\Phi_n^{(II)}(0)=1$. We observe the $r$-behaviors of $\Phi^{(II)}_1(r)$ and $\Phi^{(II)}_2(r)$ are similar to those from the two-mass-term fits, while $\Phi_3^{(II)}(r)$ has no radial nodes. In this sense, the lowest two states can be identified to be $\eta_c(1S)$ and $\eta_c(2S)$, while the third state is significantly higher and can not
 be the purely $\eta_c(3S)$ state but is likely dominated by the would-be lowest hybrid charmonium state.

If $t_{\rm min}$ decreases further to $t_{\rm min}/a_t=2,3,4$, then the correlation functions in $0^{-+}$ and $1^{--}$ channels can be fitted by four mass terms with $\chi^2/{\rm dof}$ being 2.27, 0.99 and 1.06 for $0^{-+}$, and
1.91, 1.04, and 1.05 for $1^{--}$. The fitted masses and $\Phi_n^{(II)}(0)$ of $0^{-+}$ states are listed at the bottom part of Table~\ref{tab:spec_1-II}. Although the mass $m_1$ of the lowest state is compatible with (actually slightly smaller than) the $\eta_c(1S)$ state, $m_2$ and $m_3$ are clearly different from those of the two-mass-term and three-mass-term fits.
$\Phi_n^{(II)}(r)$ functions obtained at $t_\mathrm{min}/a_t=3$ are shown in Fig.\ref{figure8}. The $r$-behavior of these states are very strange and have no radial nodes; therefore, no physical information can be inferred yet. The reason for this observation may be that, in the small $t$ range many states significantly contribute to the correlation functions including the conventional and possible charmonium-like hybrids, such that the fitted second and third state might be the admixture of them. This situation is similar to that of the four-mass-term fits for the type-I operator.

\subsection{$1^{-+}$ and $2^{-+}$ states}
 Since the quantum number $1^{-+}$ is exotic, the spectrum of this channel is much simpler than those of $0^{-+}$ and $1^{-+}$ channels. As mentioned earlier, the hybrid-like operators might couple with hybrid states predominantly in the $2^{-+}$ channel although it is permitted for conventional $D$-wave charmonia. Therefore, similar
 to the case of type-I operators, the $C_R(r,t)$ functions can be well described by the function form of three mass terms in the time window beginning from the very early time slices. The fitted results for different $t_\mathrm{min}/a_t$ values ($t_{\rm max}/a_t$ is fixed at 19) in these two channels are tabulated in Table~\ref{spec_3}. The $\chi^2/\mathrm{dof}$ values of these fits are approximately one or smaller and illustrate the goodness of the fits. We observe that the masses of the three states in each channel are in very close agreement with those of type-I operators (see Table~\ref{tab:1-+mass} and Table~\ref{tab:2-+mass}). This is exactly expected since the spectrum is independent of the operators for the same quantum number. The $\Phi^{(II)}_n(r)$ functions at $r=0$ are approximately twice as large as those of type-I operators owing to the factor of two in the definition of the type-II operator at $r=0$.

\begin{figure}
	\includegraphics[height=6cm]{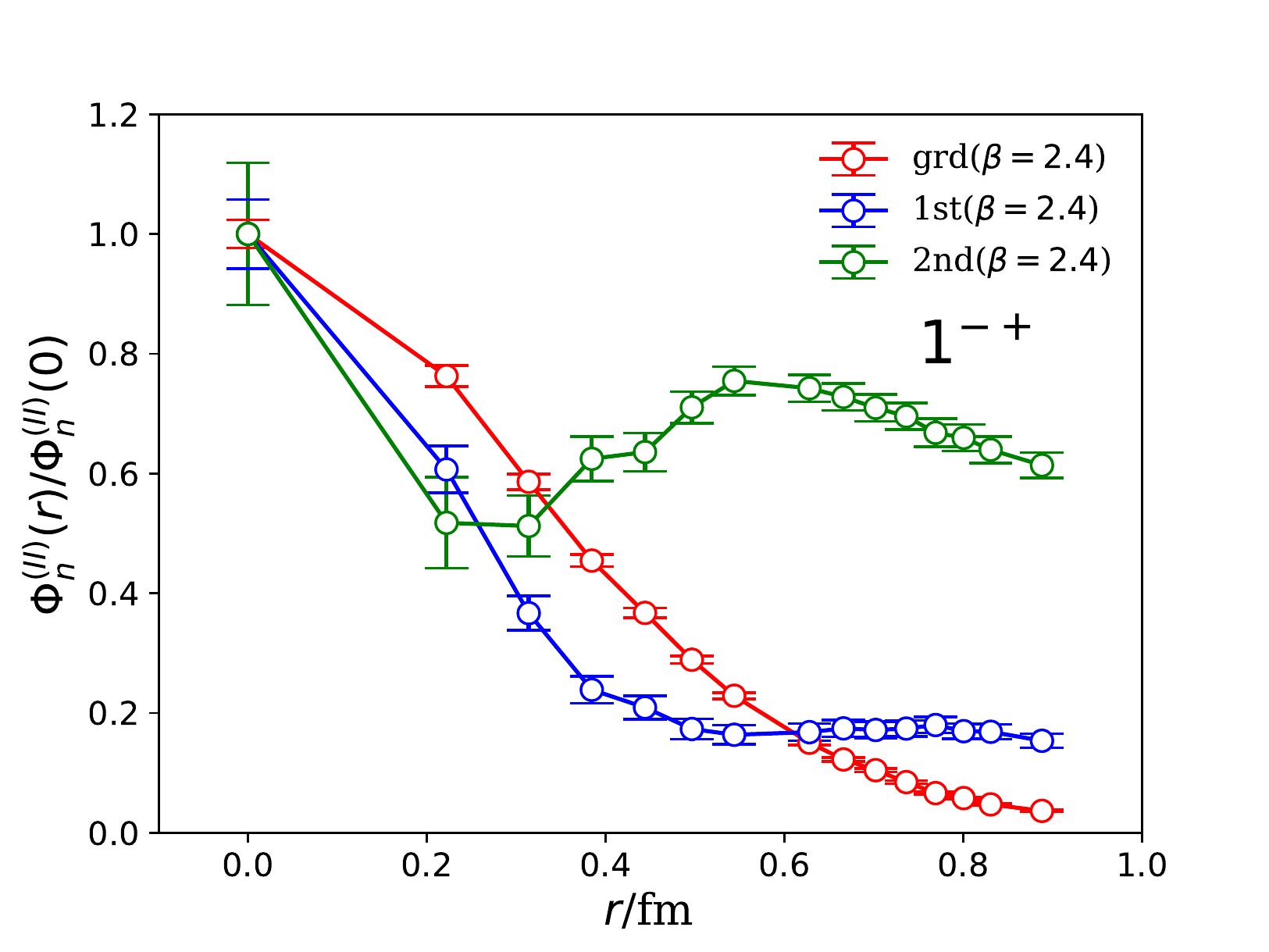}
	\includegraphics[height=6cm]{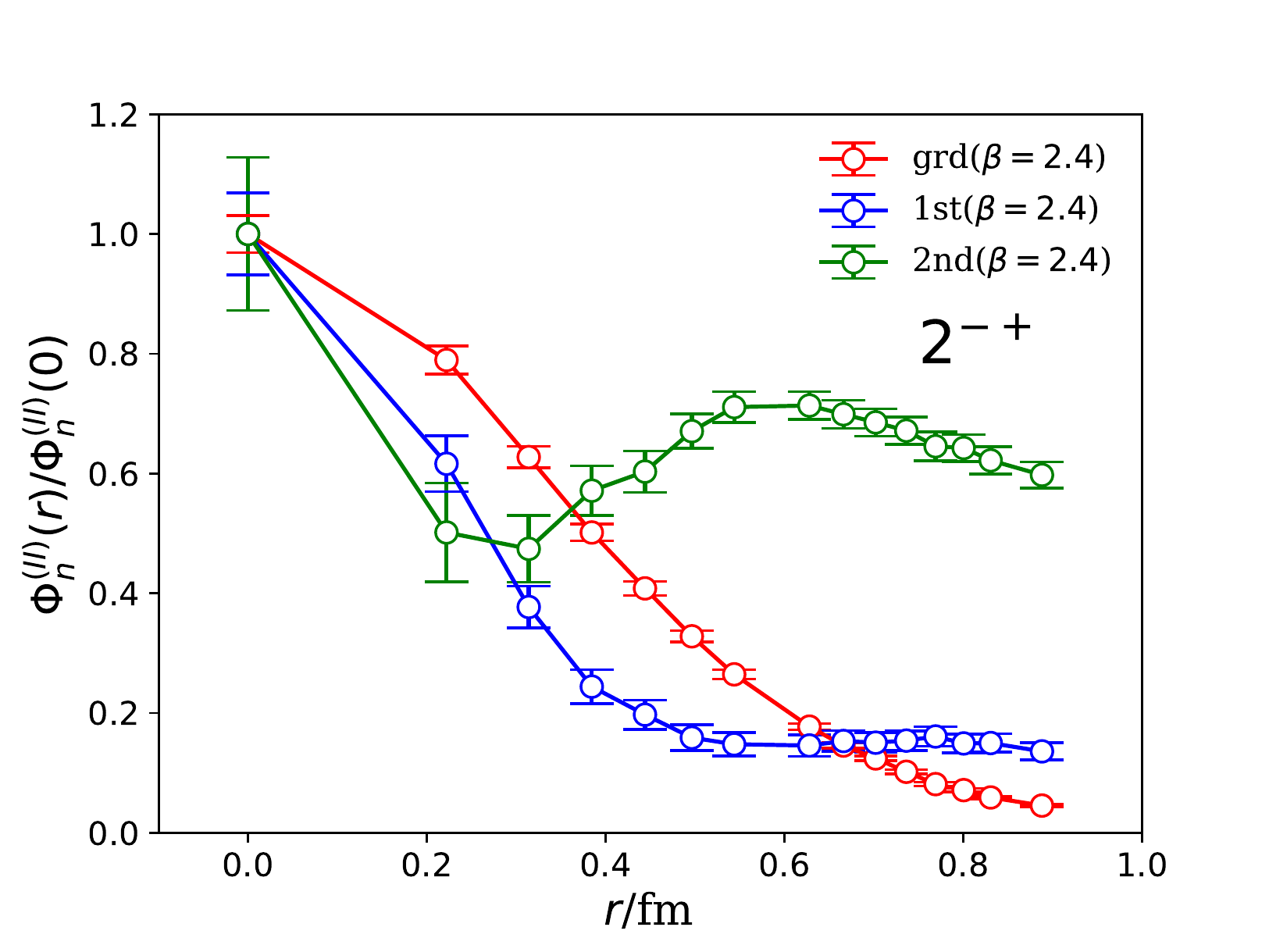}
	\caption{\label{figure9}Three-mass-term fit results of $\Phi_n^{(II)}(r)$ (normalized by $\Phi_n^{(II)}(0)=1$) of $1^{-+}$ (upper panel) and $2^{-+}$ (lower panel) states using type-II sink operators. The $x$-axis is the separation between $\bar{c}B$ and $c$ components. The results are obtained using $t_{\rm min}/a_t=3$.}
\end{figure}

 The BS wave functions $\Phi_n^{(II)}(r)$ can be derived precisely and the results at $t_{\rm min}/a_t=3$ are plotted in Fig.~\ref{figure9} (the upper panel for $1^{-+}$ and the lower one for $2^{-+}$). The $r$-behaviors are strikingly different from those of the type-I operator and there are no nodes in the $r$ direction at all. A tentative interpretation
 of this difference is that, if a charmonium-like hybrid can be considered to be a $\bar{c}-c-g$ three-body system, then its internal motion can be described through the Jacobi variables $(\mathbf{\rho},\mathbf{\lambda})$, where $\mathbf{\rho}$ is the relative displacement between $\bar{c}$ and $c$, and $\mathbf{\lambda}$ is the displacement between the gluonic degree of freedom from the center of mass of $c\bar{c}$. Thus the BS wave function from type-I operators reflect the
 internal motion with respect to $\mathbf{\lambda}$, and $\Phi_n^{(II)}(r)$ signal the projection of the full wave function $\phi(\mathbf{\rho},\mathbf{\lambda})$ onto the $r$-direction. In contrast to the BS wave function derived from the type-I operators, the $r$-behavior of $\Phi_n^{(II)}(r)$ does not exhibit the typical feature of radial excitations
 and the separation $r$ of $\bar{c}$ and $c\mathbf{B}$ components is a less meaningful variable for the pattern of excited hybrid states.
\begin{table*}[t]
	\caption{\label{spec_3} Fitted masses $m_n,n=1,2,3$ of $1^{-+}$ and $2^{-+}$ states at different $t_\mathrm{min}$ values at $\beta=2.4$, where the mass values are converted into physical units using the lattice spacings listed in Tab.~\ref{tab:lattice}. The $\chi^2/\mathrm{dof}$ value of each fit is provided to indicate the fitting quality. The $\Phi_n^{(II)}(r)$ values at $r=0$ are also listed.}
	\begin{ruledtabular}
		\begin{tabular}{cccccccc}
			&&&$1^{-+}$&&&&\\
			\hline
			$t_{\rm min}/a_t$ & $\chi^2/{\rm dof}$ & $m_1$ (GeV) & $m_2$ (GeV) & $m_3$ (GeV)
			&$\Phi_1^{(II)}(0)$      & $\Phi_2^{(II)}(0) $       & $\Phi_3^{(II)}(0)$  \\\hline
			5   &  0.494     &  4.288(16)  &  5.29(16) &  6.21(25)  & 76.3(4.4) & 70(22) & 47(27) \\
			4   &  0.489     &  4.287(9)   &  5.37(6)  &  6.86(16)  & 76.3(2.2) & 86(8)  & 43(12) \\
			3   &  0.799     &  4.290(8)   &  5.34(4)  &  6.72(7)   & 76.6(1.8) & 81(5)  & 47(6)  \\
			2   &  1.598     &  4.310(5)   &  5.53(2)  &  7.10(4)   & 82.7(1.0) & 90(2)  & 35(3)  \\
			\hline
			&&&$2^{-+}$&&&&\\
			\hline
			5   &  0.720     &  4.364(29)  &  5.20(21) &  6.09(20)  & 60.5(7.3) & 66(22) & 55(28) \\
			4   &  0.592     &  4.390(11)  &  5.48(7)  &  6.98(18)  & 68.1(2.4) & 92(9)  & 37(14) \\
			3   &  0.812     &  4.399(10)  &  5.42(5)  &  6.72(8)   & 69.2(2.1) & 79(5)  & 50(6)  \\
			2   &  1.488     &  4.414(7)   &  5.56(3)  &  7.14(4)   & 73.4(1.2) & 88(2)  & 40(3)  \\	
		\end{tabular}
	\end{ruledtabular}
\end{table*}

\section{Discussion}
In the previous sections, we presented the spectrum and BS wave functions for $1^{--},(0,1,2)^{-+}$ states by fitting the correlation functions of type-I and type-II hybrid-like operators. The results are summarized as follows:
\begin{itemize}
	\item The spectrum is independent of the operators, as expected. If we exclude the lowest two states in $0^{-+}$ and $1^{--}$ channels, which correspond to the $1S$ and $2S$ conventional charmonia, all the four channels contain states that are nearly degenerate in mass of approximately 4.4 GeV and around 5.6 GeV, respectively, which can be tentatively assigned to be the ground and first excited charmonium-like hybrid supermultiplet, respectively. Note that the mass splitting between the ground state and first excited state is approximately 1.2 GeV. This is in contrast to the
	$1S-2S$ mass splitting of conventional charmonia, which is approximately 0.6 GeV.
	
	\item The BS wave functions derived from type-I operators have a clear physical meaning. In all the four channels, the BS wave functions of the states at approximately 4.4 GeV have almost the same behavior with respect to $r$, the spatial distance between the $c\bar{c}$ component and gluonic component $\mathbf{B}$ of type-I operators. The $r$ behavior of the BS wave functions of the states at approximately 5.6 GeV are also very similar and each of them has a clear radial node at almost the same $r$. This implies that the $1^{--},(0,1,2)^{-+}$ hybrids have the same internal dynamics while the different couplings of the spin of $c\bar{c}$ and the spin of gluonic degrees of freedom result in different quantum numbers. In other words, this $r$ is a meaningful dynamical variable for charmonium-like hybrids.
	
	\item   For the states in these four channels, we also obtain the BS wave functions $\Phi_n^{(II)}(r)$ which are defined through type-II operators $O_R^{(II))}(r,t)$, with $r$ here being the spatial distance between the charm quark field $\bar{c}$ and $c\mathbf{B}$ components. In $0^{-+}$ and $1^{--}$ channels, the lowest two states in each channel have the masses compatible with those of the $1S$ and $2S$ charmonia, and the $r$-behaviors of their BS wave functions  $O_R^{(II))}(r,t)$ have the similar feature of the non-relativistic Schr\"{o}dinger wave functions of $c\bar{c}$ systems in $nS$ states. However, for higher states in the four channels, no radial nodes are observed in $\Phi_n^{(II)}(r)$. If these higher states are tentatively assigned to be hybrids, this observation might imply that
	the $r$ here is less meaningful to describe the internal motion of hybrid than the distance between the $c\bar{c}$ component and chromomagnetic field strength $\mathbf{B}$.
\end{itemize}
These results can be interpreted as follows. Since a charm quark is heavy and if the relativistic effect is not important, the $\Phi_n^{(I)}(r)$ function defined through a type-I operator may be considered the approximation of the radial wave function of a charmonium-like hybrid state to some extent. Thus, $\Phi_n^{(I)}(r)$ implies that the color octet $c\bar{c}$ pair resumes a center-of-mass motion recoiling against some additional degrees of freedom, which are necessarily gluonic in the quenched approximation. We consider the gluonic degree of freedom to be a 'constituent' gluon in the chromomagnetic mode, which functions as a color octet source and provides a potential in the non-relativistic picture. A more conceptually reasonable picture is considering the charmonium-like hybrids to be a color octet $c\bar{c}$ pair dressed by a {\it color halo} composed of gluons. In both scenarios, the binding mechanism is the strong interaction between color octets. Previous lattice studies demonstrated that for a pair of static color charge and anti-charge, their interacting potential is Cornell type~\cite{Bali:2000un},
\begin{equation}
V_D(r)=V_{0,D}-\frac{\alpha_D}{r}+\sigma_D r
\end{equation}
where the subscript $D$ indicates the color $SU(3)$ representation of the charge, $\alpha_D$ and $\sigma_D$ are proportional to the eigenvalue $C_D$ of the second order Casimir operator in the $D$ representation. This is called ``Casimir scaling". The static potential of a heavy quark-antiquark pair is usually expressed as $V_{Q\bar{Q}}(r)=V_0-4\alpha_s/3r +\sigma r$, where the string tension $\sigma$ is proportional to $C_F=4/3$. Therefore, for the color octet charge in the adjoint representation $A$ of the color $SU(3)$ with $C_A=3$, we obtain $\sigma_A=C_A/C_F \sigma=9\sigma/4$. If the excitation of a ground state hybrid is primarily along the direction from the center-of-mass of the $c\bar{c}$ to the gluonic degree of freedom, as is manifested by the wave function $\Phi_n(r)$, then the quantum number $J^{PC}$ will not change and the excitation energy will be larger than that of the charmonium system since $\sigma_A$ is 2.25 times larger. Although we cannot yet derive the precise relation of the excitation energy to $\sigma_A$, this may explain qualitatively that the mass splitting (approximately 1.2 GeV) of the ground state and first excited state hybrids is larger than the $1S-2S$ mass splitting (about 0.6 GeV) of charmonium. We observe that the ``color halo" picture can be also applied to the strangeonium-like hybrids~\cite{Ma:2020bex}.

Obviously, the ``color halo" picture we propose is clearly conceptually different from the flux-tube picture of the hybrid involving a heavy quark-anti-quark pair $Q\bar{Q}$ based on the Born-Oppenheimer approximation~\cite{Isgur:1984bm,Juge:1997nc,Juge:1999ie, Braaten:2014qka, Akbar:2020vuo}. In the Born-Oppenheimer approximation, the gluonic degrees of freedom are considered to be light and fast and distributed along the displacement of the quark-antiquark pair, whose non-trivial representation along with the spin of the gluonic degrees of freedom indicates the specific quantum number of the potential and subsequently the $J^{PC}$ of the hybrid. On the other hand, the $Q\bar{Q}$ is bound by an effective potential induced by the excitation of gluonic degrees of freedom. If the $Q\bar{Q}$ pair can be considered static color sources, the excited gluonic degrees of freedom obey the cylinder symmetry along the $Q\bar{Q}$ axis, which result in an excited static potential denoted by $\Lambda_\eta^\epsilon$, where $\Lambda=0,1, 2,\ldots$ is the projected total angular momentum of gluons with respect to the $Q\bar{Q}$ axis and is labeled as $\Sigma, \Pi, \Delta$ for $\Lambda=0,1,2$ etx., $\eta $ represents the combined parity ($P$) and the charge conjugate ($C$) of gluon excitations with $\eta=g,u$ for $P\otimes C=\pm$, respectively, and $\epsilon$ is the $P$ parity of the glue state. Therefore, the quantum number of a $Q\bar{Q}$ state with this type of potential is
\begin{equation}
P=\epsilon (-1)^{L+\Lambda+1}, C=\eta\epsilon (-1)^{L+S+\Lambda}
\end{equation}
where $\hat{L} =\hat{L}_{Q\bar{Q}}+\hat{J}_g$ with $\hat{L}_{Q\bar{Q}}$ is the orbital angular momentum of $Q\bar{Q}$ with respect to the midpoint of the $Q\bar{Q}$ axis, and $\hat{J}_g$ is the total angular momentum of gluons. The conventional Cornell-type potential between the $Q\bar{Q}$ pair is indicated by the ground $\Sigma_g^+$ potential thus the $P$ and $C$ quantum number reproduce the conventional quantum number. The lowest $1^{--}$ and $(0,1,2)^{-+}$ hybrid supermultiplet is associated with the $\Pi^+_u (L=1)$ potential such that the radial Shr\"{o}dinger equation can be expressed explicitly as
\begin{equation}\label{eq:shroedinger}
\frac{d^2}{dr^2} u(r)+2\mu[E-V_{\rm eff}(r)]u(r)=0
\end{equation}
where $r$ is the distance between $Q$ and $\bar{Q}$, $\mu$ is the reduced mass of the $Q\bar{Q}$ pair, and $u(r)$ is related to the radial wave function $\phi(r)$ by $u(r)=r\phi(r)$. The effective potential $V_{\rm eff}$ is
\begin{equation}
V_{\rm eff}=V_{Q\bar{Q}}(r)+\frac{\langle \hat{L}_{Q\bar{Q}}^2\rangle}{2\mu r^2}
\end{equation}
where $\langle \hat{L}_{Q\bar{Q}}^2\rangle=L(L+1)-2\Lambda^2+\langle \hat{J}_g^2\rangle$ and $\langle \hat{J}_g^2\rangle=2$. The eigenvalue $E$ is independent of the total spin $S$ of the $Q\bar{Q}$ pair and thereby results in the $1^{--}$ and $(0,1,2)^{-+}$ supermultiplet. When the parameters of $V_{Q\bar{Q}}$ are set phenomenologically or from the lattice QCD results, the Shr\"{o}dinger equation in Eq.~(\ref{eq:shroedinger}) can be solved and the mass spectrum can be obtained. In Ref.~\cite{Braaten:2014qka}, the mass splitting between the lowest $\Pi_u^+$ multiplet and its first radial excitation is determined to be approximately 350 MeV for charmonium-like hybrids and 207 MeV for bottomium-like hybrids, which are significantly smaller than 1.2 GeV obtained in this paper. Moreover, the wave function with respect to the distance between the $Q\bar{Q}$ pair should behave like a $P$-wave one, which is clearly different from the behaviors of the wave functions $\Phi_n^{(II)}(r)$. In other words, our results do not support the flux-tube description of heavy-quarkonium-like hybrids in the Born-Oppenheimer approximation. It should be emphasized that even though the interpretation of the wave functions can be debatable, the pattern of the spectrum should be
solid and model-independent since it is derived directly from the lattice QCD calculation.

The ``color halo" picture can have physical consequences. Based on the discussion above, the $c\bar{c}$ pair and gluonic component are bound through the potential of color octets. This binding can be easily broken by the excitation of gluons such that the $c\bar{c}$ component is neutralized in color and is emitted as a conventional charmonium, while the gluonic
component is hadronized into light hadrons. That is, the decay modes of a charmonium state plus light hadrons can be important for the decay of a charmonium-like hybrid. This decay property seems compatible with the decay pattern of $Y(4260)$ (now named $\psi(4230)$ {\it aka} $Y(4230)$ by the PDG), which is occasionally assigned to be a candidate for the $1^{--}$ charmonium-like hybrid. $Y(4260)$ was first observed in the invariant mass spectrum of $J/\psi \pi^+\pi^-$. Over the past several years, BESIII has observed structures at the center-of-mass of 4.22-4.23 GeV of the final states $J/\psi\pi\pi, \chi_{c0}\omega, h_c\pi\pi, \psi(3686)\pi\pi $ in $e^+e^-$ annihilations~\cite{Yuan:2018inv}, which can be the previous $Y(4260)$ if they come from the same state. The cross sections of these processes are comparable at the peak positions and can be understood by the hybrid assignment of $Y(4260)$. Note that the $\bar{c}c$ pair in the $1^{--}$ hybrid is a spin singlet, the decay modes involving $h_c$ and $\eta_c$ should be preferable to those involving $J/\psi, \chi_{c0}$ and $\psi(3686)$ owing to the suppression of the spin flipping of charm quarks. However, the $h_c\pi^+\pi^-$ decay mode, in which $h_c$ and $\pi^+\pi^-$ are in relative $P$-wave, is suppressed by the centrifugal potential barrier in contrast to the other channels in which the charmonium and light hadrons are in $S$-wave. The two effects compensate and result in comparable cross sections. On the other hand, the $1^{--}$ states $Y(4360)$ and $Y(4660)$, if they do exist, are disfavored to be the radial excitations of $1^{--}$ charmonium hybrids, since their masses are much lower than the predicted value in this work.

The $(0,1,2)^{-+}$ charmonium-like hybrids can be also searched in the decay modes involving a charmonium state, among which the final state $\chi_{c0,1,2}\eta$ may be important, since no suppression occurs from the spin-flipping of heavy quarks and the centrifugal potential barriers. The disadvantage of these modes is that $\eta$ only has a small fraction of the flavor singlet component, but the QCD anomaly may enhance its production to some extent if it couples to two gluons. Other modes, such as $J/\psi \omega$ and $J/\psi \phi$ (in $P$-wave), are worth considering, which are similar to the $J/\psi \pi^+\pi^-$ and $\chi_{c0} \omega$ decay modes of the $1^{--}$ case. Since these states are heavy and cannot be observed directly in the $e^+e^-$ annihilation, BelleII and LHCb may implement a mission to search for them in $B$ meson decays.

\section{Summary}
The internal structures of the $J^{PC}=1^{--}, (0,1,2)^{-+}$ charmonium-like
hybrids are investigated for the first time through their BS wave functions from lattice QCD in the quenched
approximation, where the wave functions $\Phi_n^{(I)}(r)$ are defined by the state-to-vacuum matrix elements of spatially extended hybrid-like operators (type-I operators) with the color octet $\bar{c}c$ component separated from the chromomagnetic operator by a spatial distance $r$. After singling out the conventional $\bar{c}c$ states in the $0^{-+}$ and $1^{--}$ channels, we confirmed the existence of a $1^{--}$ and $(0,1,2)^{-+}$ supermultiplet of nearly degenerate masses of approximately $4.3-4.6$ GeV and similar BS wave functions $\Phi_1(r)$ without nodes in the $r$ direction. The first excited hybrid states also compose a supermultiplet with masses of approximately 5.7 GeV, and their BS wave functions $\Phi_2(r)$ are almost the same and have one node. We checked these results by using type-II operators that 
the charm quark field $\bar{c}$ is separated from the $c\mathbf{B}$ component by a spatial distance. While the spectra from the two types of operators are consistent with each other, the wave functions $\Phi^{(II)}_n(r)$ do not have a nodal structure with respect to the distance between $\bar{c}$ and $c\mathbf{B}$. 
These observations imply that $r$ can be a significant dynamical variable for charmonium-like hybrids. The spectrum and information from the wave functions obtained in this study do not support the flux-tube description of heavy quarkonium-like hybrids in the Born-Oppenheimer approximation. Instead, we propose a ``color-halo" scenario for the internal structure of the charmonium-like hybrids in which a relatively compact color octet $\bar{c}c$ pair is surrounded by gluonic degrees of freedom. Thus, the decay modes of a charmonium plus light hadrons are important for charmonium-like hybrids. Finally, we advocate for LHCb and BelleII to search for these charmonium-like hybrids in $B$ decays. Specifically, $(0,1,2)^{-+}$ hybrids can be searched in $\chi_{c0,1,2}\eta$ and $J/\psi\omega (\phi)$ systems.

\section*{ACKNOWLEDGEMENTS}
This work is supported by the National Key Research and Development Program of China
(No.2017YFB0203202) and the Strategic Priority Research Program of Chinese Academy of Sciences (XDB34030302 and No.XDC01040100), the National Natural Science Foundation of China (NNSFC) under Grants No. 11935017, No. 11575196, No. 11775229, No. 12075253, and No. 12070131001 (CRC 110 by DFG and NSFC). The numerical calculations are carried out on Tianhe-1A at the National Supercomputer Center (NSCC) in Tianjin and the GPU cluster at IHEP. Y.C. is also supported by the CAS Center for Excellence in Particle Physics (CCEPP). W.S. thanks the support of DOE under contract DE-AC05-06OR23177.

\bibliography{ref}

\end{document}